%% file: neurips_2024.tex
\documentclass{article}

\usepackage[final]{neurips_2024}

\input{math_commands.tex}

\usepackage{shortcuts}

\usepackage{hyperref, xcolor}
\usepackage{natbib}
\definecolor{linkcolor}{RGB}{83,83,182}
\hypersetup{
    colorlinks=true,
    citecolor=linkcolor,
    linkcolor=linkcolor
}
\usepackage{url}

\usepackage{algorithm,algorithmic}

\theoremstyle{plain}

\title{Unmixing Noise from Hawkes Process to\\Model Learned Physiological Events}

\author{%
  Guillaume Staerman \\
  Université Paris-Saclay, Inria, CEA\\
 Palaiseau, France \\
  \texttt{guillaume.staerman@inria.fr} \\
  \And
  Virginie Loison \\
  Université Paris-Saclay, Inria, CEA\\
 Palaiseau, France \\
  \texttt{virginie.loison@inria.fr} \\
  \AND
  Thomas Moreau \\
  Université Paris-Saclay, Inria, CEA\\
 Palaiseau, France \\
  \texttt{thomas.moreau@inria.fr} \\
}

\begin{document}

\maketitle

\begin{abstract}
Physiological signal analysis often involves identifying events crucial to understanding biological dynamics. Traditional methods rely on handcrafted procedures or supervised learning, presenting challenges such as expert dependence, lack of robustness, and the need for extensive labeled data. Data-driven methods like Convolutional Dictionary Learning (CDL) offer an alternative but tend to produce spurious detections. This work introduces UNHaP (Unmix Noise from Hawkes Processes), a novel approach addressing the joint learning of temporal structures in events and the removal of spurious detections. Leveraging marked Hawkes processes, UNHaP distinguishes between events of interest and spurious ones. By treating the event detection output as a mixture of structured and unstructured events, UNHaP efficiently unmixes these processes and estimates their parameters. This approach significantly enhances the understanding of event distributions while minimizing false detection rates.
\end{abstract}

\section{Introduction}

The analysis of physiological signals often boils down to identifying events of interest. Typical examples are with electrocardiography (ECG), where the detection of the QRS complex --\emph{a.k.a.} the heartbeat-- is a fundamental step to characterize the status of the cardiovascular system, with biomarkers like the heart rate (HR; \citealp{berkaya2018survey}) and heart rate variability (HRV;~\citealp{Luz2016}).
Another example is the identification of steps in inertial measurement unit recordings, which is a crucial feature in classifying pathological gait anomalies \citep{cimolin2014summary}.

To automatize the event detection step, several approaches have been proposed.
In most physiological signal processing applications, events are detected with handcrafted procedures based on signal processing techniques.
For instance, the QRS complexes or the steps are identified using peak detection algorithms~\citep{Pan1985} or wavelet-based approaches~\citep{Martinez2004}.
%
While these algorithms perform well, they require large domain expertise, and their parameters tend to be sensible to the acquisition protocol.
Data-driven approaches have also been proposed, using supervised deep learning~\citep{xiang2018automatic, Craik2019}.
These approaches demonstrate excellent performance on particular tasks.
Yet, they require large labeled datasets.
Another data-driven approach is unsupervised learning to extract repeating patterns, such as the convolutional dictionary learning (CDL) algorithm~\citep{Grosse2007}.
These methods aim to represent events through their prototypical patterns, which are directly learned from the data.
While these solutions can be applied without domain expertise, they tend to detect more spurious events.

To reach satisfactory results, all these methods require post-processing steps to filter out spurious events.
Developing and characterizing these extra steps is a tedious task, requiring domain expertise and time.
In this paper, we propose a novel automatized framework to filter out spurious events based on their temporal distribution and the event detection confidence.
A key observation for all event detection methods is that each event is detected independently, with an estimated confidence in the event detection.
However, in most cases, the events are distributed with an informative temporal structure. For instance, the inter-heartbeat interval is around one second for a normal ECG.
We propose to classify detected events between spurious and structured ones, by jointly learning the temporal structure of the events and filtering out spurious event detection based on the distribution of confidence levels.

To model the events' temporal distribution, we rely on Hawkes processes (HP; ~\citealp{Hawkes1971}), a classical type of point process (PP) to model past events' influence on future events.
Recent works have proposed novel inference techniques adapted to physiological events' distribution~\citep{Allain2022, staerman2023fadin}.
%
Yet, these models can't account for the confidence associated with the event detection. This confidence could be quantified by marks. These models therefore need to be extended to deal with marked PP~\citep{daley2003introduction} and to integrate marks in the intensity function.

In addition, inference with these models only works when all events come from the same process.
In our context, a mixture of spurious events from a noise process and structured events is observed, and direct inference gives uninformative biased results.
Mixtures of Hawkes processes have been considered in the literature either to cluster events \citep{liu2019latent,yang2013mixture} or sequences of events~\citep{xu2017dirichlet}. They rely on feature-based mixture models \citep{li2013dyadic,yang2013mixture,du2015dirichlet} or associate a Dirichlet process to classical Hawkes models \citep{Blei2006}. While these approaches are tailored to find different auto-excitation patterns, they are not designed to unmix noise and uninformative events from structured ones.

%

\noindent \textbf{Contributions.}
To jointly model the temporal distribution of events and remove spurious events, we propose a novel method named UNHaP to Unmix Noise from Hawkes Processes.
In our model, the output of the event detection algorithm is treated as a mixture of events of interest with a Hawkes process structure and spurious events that are not of interest, distributed as a Poisson process.
UNHaP aims to learn to distinguish between these two distinct processes to select properly structured events and discard the spurious ones.
Based on the FaDIn framework~\citep{staerman2023fadin}, we propose an efficient algorithm to jointly unmix these events and estimate the parameters of the Hawkes process. We illustrate the benefits of using our unmixing models rather than the traditional Hawkes process models with real-world ECG and gait data.

\section{Background on Marked Hawkes Processes}


A multivariate marked Hawkes process (MMHP) is a self-exciting point process that models the occurrence of events in time, where each event is associated with supplementary information, referred to as the ``\emph{mark}'' of the event. The mark may or may not integrate the event type in the literature. Throughout this paper, we separate the event type from the mark and consider continuous marks belonging to $\mathbb{R}$. We here give our notation and basic information about MMHP and refer the reader to \cite[Sec. 6]{daley2003introduction} for a detailed account of these processes.

\noindent \textbf{Counting processes.}
Let $\mathscr{F}_{T}$ be a set of observed marked events including $D$ types such that  for each $i\in \intervalleEntier{1}{D}$ we have  {\small $\mathscr{F}_{T}^i = \big \{(t_n^i, \kappa_n^i): \;  \kappa_n^i \in \mathcal{K}, \;  t_n^i \in [0, T]\big \} $} with $t_n^i$ the time where the $n$-th event of type $i$ occurs and $\kappa_n^i$ its associated mark.
We denote by $\mathbf{N}_i$ the random counting measure defined on $[0, T] \times \mathbb{R}_+$, such that $\mathbf{N}_i(\mathrm{d}t, \mathrm{d}\kappa) = \sum_{n=1}^{\infty}\delta_{(t_n^i, \kappa_n^i)} (\mathrm{d}t , \mathrm{d} \kappa ),$ where $t$ and $\kappa$ represent respectively the time and the mark, and $T\in \mathbb{R}_+$ is the stopping time.
Without limitations, the set of marks is assumed to be any compact set $\mathcal{K} \subset \mathbb{R}_+$.
From this measure, we can define the marginal time arrival process, also called ground process, as $N_{i}(T) = \int_{[0, T]\times \mathbb{R}_+}\mathbf{N}_i(\mathrm{d}t, \mathrm{d}\kappa)=\sum\limits_{n \geq 1} \1[t_n^i \leq T]$.

\noindent \textbf{Intensity function.}
The behavior of a MMHP can be described by its intensity function.
Conditionally to observed events, it describes the instantaneous event rate at any given point in time.
Given a MMHP and a set of observation $\mathscr{F}_T=\{ \mathscr{F}_T^i\}_{i=1}^D$, each ground process $N_{i}$ is described by the following conditional ground intensity function
\begin{align*}
    \lambda_{g_i}(t | \mathscr{F}_t) \nonumber = \mu_i + \sum_{j=1}^D  \int_{[0, t) \times \mathcal{K}}  h_{ij}(t-u, \kappa)~\mathbf{N}_j (\mathrm{d}u, \mathrm{d \kappa}),
\end{align*}
where $\mu_i$ is the baseline rate and $h_{ij}:  \mathbb{R}_+ \times \mathcal{K} \rightarrow \mathbb{R}_+$ is the triggering or kernel function, quantifying the influence of the $j$-th process' past events onto the $i$-th process' future events.
The ground intensity quantifies the time probability of future events, taking into account the marks of previous events.
In the following, we consider independent probability for the marks \citep{daley2003introduction}, assuming a factorized form for the kernel $h_{ij}(t, \kappa) = \phi_{ij}(t) \omega_{ij}(\kappa)$.
This leads to%
{\small%
\begin{align*}
    \lambda_{g_i}(t | \mathscr{F}_t) \nonumber &= \mu_i + \sum_{j=1}^D  \int_{[0, t) \times \mathcal{K}}  \omega_{ij}(\kappa)~\phi_{ij}(t - u) ~\mathbf{N}_j (\mathrm{d}u \times \mathrm{d \kappa}) \nonumber  = \mu_i + \sum_{j=1}^D  \sum_{n, t_n^j < t} \omega_{ij}(\kappa_n^j)~\phi_{ij}(t - t_n^j),
\end{align*}}%
\looseness=-1
with $\omega_{ij}: \mathcal{K} \rightarrow \mathbb{R}_+$, $\phi_{ij}: \mathbb{R}_+ \rightarrow \mathbb{R}_+$ such that $\int_{0}^{\infty} \phi_{ij}(t)\mathrm{d}t < 1$ and $\int_{\mathcal{K}} \omega_{ij}(\kappa)\mathrm{d}\kappa < 1$. These conditions ensure the stability of such processes.  The function $\omega_{ij}(\cdot)$ weights the probability that a future event occurs depending on the past events' marks. Assuming a collection $\{f_{i}: \mathcal{K} \rightarrow \mathbb{R}_+\}_{i=1}^{D}$ of density functions, we define the joint intensity function as $\lambda_i (t, \kappa) = \lambda_{g_i}(t | \mathscr{F}_t)~ f_i(\kappa),$
where the ground process depends on the mark distribution reflected by $f_i$ and the distribution of the influence of the mark described by $\omega_{ij}$.


\noindent \textbf{ERM-based inference.}
Inference for MMHP is usually performed using the log-likelihood to align the model with the observed data~\citep{daley2003introduction,bacry2015hawkes}.
While this can be efficient for Markovian kernels, it becomes computationally expensive for more general ones~\citep{staerman2023fadin}.
In this paper, we instead resort to the Empirical Risk Minimization (ERM)-inspired least squares loss (refer to Eq. (II.4) in \citealp[Chapter~2]{bompaire:tel-02316143}).
The goal is to  minimize
%
\begin{equation}\label{eq:l2continuous}
    \mathcal{L}\pars{\boldsymbol{\theta}, \mathscr{F}_T} = \sum_{i=1}^{D}  \pars{\int_{0}^{T} \hspace{-0.2cm} \int_{\mathcal{K}}\lambda_{i}(s, \kappa; \boldsymbol{\theta})^2~\mathrm{d}\kappa\mathrm{d}s - 2 \sum_{(t_n^{i}, \kappa_n^i) \in \mathscr{F}_T^i} \lambda_{i}\pars{t_n^{i}, \kappa_n^i; \boldsymbol{\theta}}} ,
\end{equation}
%
where $\boldsymbol{\theta}=\{\mu_i, \phi_{ij}, \omega_{ij} \}_{i=1}^{D}$.
This loss function corresponds to the empirical approximation of the expected risk incurred by the model measured by $\|\lambda(\boldsymbol{\theta}) - \lambda^*\|_2$, with $\lambda^*$ the true intensity function.
It is  more efficient to compute than the log-likelihood, especially for general parametric kernels~\citep{staerman2023fadin}.


\section{Unmixing Noise from Hawkes Process}

\noindent \textbf{Problem statement.} We consider a set of observed events $\mathscr{F}_T =\big\{ e^i_n=(t^{i}_n, \kappa^{i}_n), \; 1\leq n \leq N_{i}(T)\}_{i=1}^{D}$ with events originating from two independent processes.
We denote $\mathscr{F}_{T,k}=\big\{e_n^{i,k} = (t^{i,k}_n, \kappa^{i,k}_n); 1 \le n \le  N^k_{i}(T) \big \}_{i=1}^{D}$ these two processes such that $\mathscr{F}_T = \mathscr{F}_{T,0} \cup \mathscr{F}_{T,1}$.
We consider the case where $\mathscr{F}_{T,0}$ is a homogeneous marked Poisson process --representing spurious event detections-- and $\mathscr{F}_{T,1}$ is a MMHP --for structured events.
This problem is a denoising problem, where spurious events are considered as noise that should be discarded for the application.

Our goal is to unmix these two processes, \ie{} to associate each event $e^i_n \in \mathscr{F}_T$ with a label $Y^i_n \in \{0, 1\}$ such that $Y_n^i =1$, if $e^i_n$ originates from $\mathscr{F}_{T,1}$.
This task amounts to binary classification for the events.
However, the main difficulty lies in that the labels are unknown, and the events are not independent.
To cope with the lack of labels, we propose to leverage the temporal MMHP structure of $\mathscr{F}_{T,1}$ to characterize structured events, assigning events with this process if they are plausible according to the MMHP model.
This is an arduous assignment problem, which we address using a variational inference approach and a mean-field relaxation.
This procedure allows us to jointly estimate the parameters of the processes while unmixing the events, see \autoref{fig:illustration}.

\definecolor{structured}{RGB}{0, 128, 0}
\definecolor{spurious}{RGB}{255, 0, 0}
\definecolor{observed}{RGB}{0, 0, 255}
\definecolor{intensity}{RGB}{107, 107, 107}
\begin{figure}[t]
    \includegraphics[width=\textwidth]{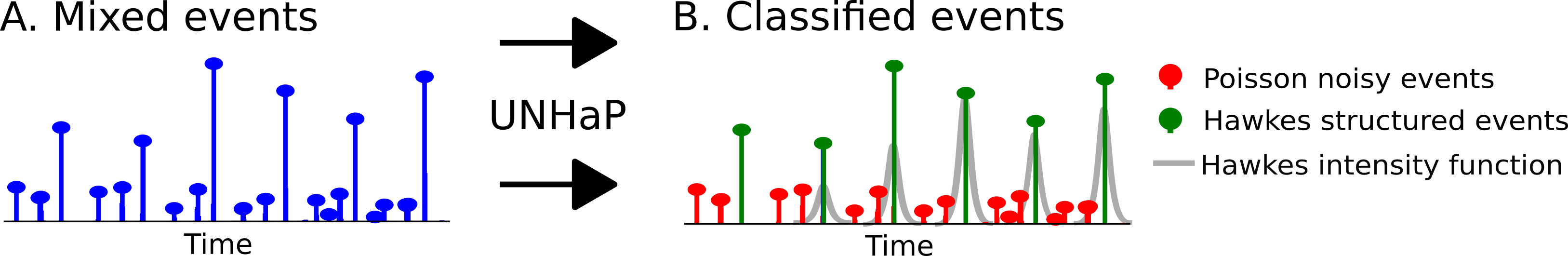}
    \caption{
        {\bf Illustration of the UNHaP framework.~}
        The goal of UNHaP is to distinguish between structured events (\textcolor{structured}{\bf green}) and spurious ones (\textcolor{spurious}{\bf red}) by identifying the structure of the MMHP (\textcolor{intensity}{\bf grey}) from the observed events (\textcolor{observed}{\bf blue}).}
        \label{fig:illustration}
\end{figure}




\noindent \textbf{Latent variables and risk function.} Unmixing noise from MMHP events amounts to a binary classification task, where the underlying structure of the events allows to discriminate between the two classes and has to be inferred.
Our goal is thus to infer the value of latent variables $Y_n^i$ for each event such that $Y_n^i =1$ if the $n$-th event of the $i$-th type is generated by $\mathscr{F}_{T,1}$ while $Y_n^i=0$ if it is generated by $\mathscr{F}_{T,0}$.


When these latent variables are known, it is possible to write the intensity functions of both processes from the observed events $\mathscr{F}_T$.
Spurious events from $\mathscr{F}_{T, 0}$ are distributed following a marked Poisson process with intensity $\lambda_i^0(t, \kappa; \boldsymbol{\theta}_0) = \tilde{\mu}_i f_i^0(\kappa)$ such that $\tilde{\mu}_i \in \mathbb{R}_+$, $f_i^0: \mathcal{K} \rightarrow  \mathbb{R}_+$, $\int_{\mathcal{K}} f_i^0(\kappa)~\mathrm{d}\kappa = 1$ and $\boldsymbol{\theta}_0=\{\tilde{\mu}_i \}_{i=1}^{D}$.
Non-spurious events follow a MMHP whose intensity, denoted $\lambda_i^1(t, \kappa; \boldsymbol{\theta}_1)$, can be derived from the observed events only.
We have, for $t \in [0, T]$,
\begin{align*}
    \lambda_i^1(t, \kappa; \boldsymbol{\theta}_1) = \Big(\mu_i + \sum_{j=1}^{D}\sum_{t_n^j < t} Y_n^j\phi_{ij}(t - t_n^j; \eta_{ij})~\omega_{ij}(\kappa_n^j)\Big)f_i^1(\kappa),
\end{align*}
where $\phi_{ij}$ is a parametric kernel parametrized by $\eta_{ij}$ and $\boldsymbol{\theta}_1=\{\mu_i, \eta_{ij}\}_{i,j=1}^{D}$.
An important remark is that the intensity function depends only on past events from  $\mathscr{F}_{T, 1}$.
This is where our model differs from classical MHHP models, as it is necessary to select the right events to be able to compute the intensity function.

\looseness=-1
Conditioned on the latent variables $\{Y_n^i\}$, both processes are independent.
The risk for the parameters $\boldsymbol{\theta}$ is thus the sum of the least square loss, defined in (\ref{eq:l2continuous}), for each process, \emph{i.e.}, $\mathcal{L}(\boldsymbol{\theta};\mathscr{F}_{T}) =\mathcal{L}(\boldsymbol{\theta}_0; \mathscr{F}_{T,0})+\mathcal{L}(\boldsymbol{\theta}_1; \mathscr{F}_{T,1})$.
The complete loss, assuming $\mathcal{Y}_T=\{Y_n^i\}_{i, n}$ are observed, can thus be written as $\mathcal{L}\pars{\boldsymbol{\theta}; \mathcal{Y}_T, \mathscr{F}_T} = \sum_{i=1}^D\mathcal{L}^i\pars{\boldsymbol{\theta}; \mathcal{Y}_T, \mathscr{F}_T}$, where
\begin{equation}\label{eq:loss2}
    \begin{split}
        \mathcal{L}^i\pars{\boldsymbol{\theta}; \mathcal{Y}_T, \mathscr{F}_T }  = & \int_{0}^{T} \hspace{-0.2cm} \int_{\mathcal{K}}
        \lambda_i^0(t, \kappa; \boldsymbol{\theta}_0)^2~\mathrm{d}\kappa\mathrm{d}t  + \int_{0}^{T} \hspace{-0.2cm} \int_{\mathcal{K}}\lambda_i^1(t, \kappa; \boldsymbol{\theta}_1)^2~\mathrm{d}\kappa\mathrm{d}t
        \\& \quad- 2 \sum_{ e_n^i \in \mathscr{F}_T^i} (1-Y_n^i)\lambda_i^0(t_n^i, \kappa_n^i; \boldsymbol{\theta}_0) + Y_n^i\lambda_i^1 (t_n^i, \kappa_n^i; \boldsymbol{\theta}_1).
    \end{split}
\end{equation}
If $\boldsymbol{\lambda}_i^{0}$ and $\boldsymbol{\lambda}_i^{1}$ are the true intensity functions of the underlying processes, then we have $\mathbb E_{\mathscr{F}_T}[\mathcal{L}^i\pars{\boldsymbol{\theta}; \mathcal{Y}_T, \mathscr{F}_T}] = \|\lambda_i^0(\boldsymbol{\theta_0}) - \boldsymbol{\lambda}_i^0\|_2^2 + \|\lambda_i^1(\boldsymbol{\theta_1}) - \boldsymbol{\lambda}_i^1\|_2^2 - C$ where $C$ is a constant in $\boldsymbol{\theta}$.
This loss $\mathcal{L}\pars{\boldsymbol{\theta}; \mathcal{Y}_T, \mathscr{F}_T}$ is thus the empirical risk of the model for a given set of observed events and an assignment $\{Y_n^i\}$, and the model's parameters can be inferred by minimizing it.

\noindent \textbf{Mean-field-based Variational Inference.} %
The goal of our procedure is also to infer the collection of $\{Y_n^i\}$.
The classical procedure to solve such latent factor estimation with probabilistic models is to resort to the Expectation-Maximization (EM) algorithm.
This algorithm allows the iterative refinement of the $\boldsymbol{\theta}$'s estimate by maximizing the likelihood marginalized over the latent factors $Y_n^i$.
This requires being able to compute the marginalized likelihood or at least estimate it with Monte Carlo sampling.
But this step is not possible with the assignment variable $Y_n^i$ due to the complex dependency structure between the various $Y_n^i$ imposed by the Hawkes process structure.

\looseness=-1
To alleviate this challenge, we propose to resort to a mean-field approximation with independent variables for each event. Specifically, we perform the following approximation
\begin{equation}\label{eq:mean-field}
  p(\mathbf{Y}; \mathscr F_T)= \prod_{i=1}^D p(Y^i; \mathscr F_T^i)
  \approx  \prod_{i=1}^D\prod_{n=1}^{N_T^i} q(Y_n^i; \rho_n^i),
\end{equation}
\vskip-1em%
\looseness=-1
where $q(Y; \rho)$ is a univariate Bernoulli distribution with parameter $\rho$.
The parameter $\rho_n^i$ is the probability that $Y_n^i = 1$.
It corresponds to a relaxation of the assignment variable $Y_n^i \in \{0, 1\}$ to the interval $[0, 1]$.
This relaxation allows us to compute the expected risk of the model with respect to the latent variables.
Therefore, we have $\bar{\mathcal{L}} \pars{ \boldsymbol{\rho}, \boldsymbol{\theta}; \mathscr{F}_T} = \mathbb{E}_{\mathbf{Y}} \left[ \mathcal{L}\pars{\boldsymbol{\theta}; \mathcal{Y}_T, \mathscr{F}_{T}}\right]= \sum_{i=1}^D \bar{\mathcal{L}}^i\pars{ \boldsymbol{\rho}, \boldsymbol{\theta}; \mathscr{F}_T}$
with
\begin{equation}\label{eq:loss}
\begin{split}
    \bar{\mathcal{L}}^i\pars{\boldsymbol{\theta}, \boldsymbol{\rho};  \mathscr{F}_T} = &\int_{0}^{T}\hspace{-0.2cm} \int_{\mathcal{K}}
    \lambda_i^0(t, \kappa)^2~\mathrm{d}\kappa\mathrm{d}t +\int_{0}^{T} \hspace{-0.2cm}\int_{\mathcal{K}}\bar\lambda_i^1(t, \kappa)^2~\mathrm{d}\kappa\mathrm{d}t
    + \boldsymbol{C}(\boldsymbol{\rho})\\
     & - 2 \sum_{n, t^i_n \in \mathscr{F}_T^i} \bigg( (1-\rho^i_n)\lambda_i^0\pars{t_n^i, \kappa_n^i} + \rho_n^i\bar\lambda^1\pars{t_n^i, \kappa_n^i} \bigg),
\end{split}
\end{equation}
where $\boldsymbol{\rho}=\{ \rho_n^i\}$, $\boldsymbol{C}(\boldsymbol{\rho}) = \sum_{j=1}^{D}\int_0^T \sum_{n, t_n^j<t} \rho_n^j(1 - \rho_n^j) ~\omega_{ij}(\kappa_n^j)^2\phi_{ij}(t-t_n^j)^2 \mathrm{d}t$ and {\small
$
    \bar\lambda_i^1(t, \kappa; \boldsymbol{\theta}_1) = \Big(\mu_i + \sum_{j=1}^{D}\sum_{t_n^j<t} \rho_n^j\phi_{ij}(t - t_n^j; \eta_{ij})\omega_{ij}(\kappa_n^j) \Big)f^1_i(\kappa)
$} corresponds to $\lambda^1_i$ where $\boldsymbol{Y}$ has been replaced by $\boldsymbol{\rho}$.
Here, we can replace $Y_n^i$ by its expectation $\rho_n^i$ in the integral of the squared intensity as $\mathbb{E}[Y^i_nY^i_l] = \rho_n^i\rho_l^i$ for the distribution $q$.
However, this is not true for $\mathbb{E}[(Y_n^i)^2]$ which is equal to $\rho_n^i$ and not $(\rho_n^i)^2$.
$\boldsymbol{C}(\boldsymbol{\rho})$ corrects this discrepancy.
Note that $\bar{\mathcal L}$ can also be seen as a relaxation of the assignment problem with continuous variables $\rho_n^i$.

Based on this mean-field approximation, we propose a variant of the classification EM algorithm (CEM; \citealp{celeux1992classification}) summarized in~\autoref{algo}.
The {\bf E}-step consists in minimizing $\bar{\mathcal{L}} \pars{ \boldsymbol{\rho}, \boldsymbol{\theta}^{\ell-1}; \mathscr{F}_T}$ w.r.t. the latent parameters $\boldsymbol{\rho}$.
The {\bf C}-step assigns each event to the corresponding class $\{0, 1\}$ by setting $Y_n^{i,(\ell)}=\mathbb{I}\{\rho_n^{i,(\ell)} > 1/2 \}$.
The {\bf M}-step amounts to minimizing $\mathcal{L}(\boldsymbol{\theta};  \mathcal{Y}_T, \mathscr{F}_T)$ w.r.t. $\boldsymbol{\theta}$.
Repeating these steps yields an estimation of the parameter $\boldsymbol{\theta}$, encoding the structure of the events, as well as the assignment $Y_n^i$ of each event $e_n^i$ to one of the two processes.
This procedure constitutes the core of the UNHaP unmixing procedure.
In addition to this variational procedure, fast and efficient inference in UNHaP relies on several key points described below.

\setlength{\textfloatsep}{15pt}
\begin{algorithm}[t]
    \begin{algorithmic} 
    \INPUT  Set of events $\mathscr F_T$.\\
    \noindent \hspace{-0.35cm}\textbf{initialization} $ \boldsymbol{\rho}^{(0)} \overset{i.i.d.}{\sim}  q(1/2)$, $\boldsymbol{\theta}^{(0)}$ initialized with Moments Matching.\\[.3em]
    \FOR{$\ell$=1, \ldots $n_{\mathrm{iter}}$}
    \STATE \textbf{(E-step)}  $\boldsymbol{\rho}^{(\ell)} = \underset{\boldsymbol{\rho}}{\text{argmin}} ~\sum_{i=1}^D \bar{\mathcal{L}}_{\mathcal{G}}^i(\boldsymbol{\rho};\boldsymbol{\theta}^{(\ell-1)}, \mathscr{F}_T)$


    \STATE \textbf{(C-step)} Assign the events by computing $\mathcal{Y}_T^{(\ell)} = \big \{Y_n^{i, (\ell)}=\mathbb{I}\{\rho_n^{i, (\ell)} > 1/2 \} \big\}_{i,n}$.

    \vspace{.5em}

    \STATE \textbf{(M-step)} $\boldsymbol{\theta}^{(\ell)} = \underset{\boldsymbol{\theta}}{\text{argmin}} ~\mathcal{L}_{\mathcal{G}}(\boldsymbol{\theta}; \mathcal{Y}_T^{(\ell)}, \mathscr{F}_T)$ initialized $\boldsymbol{\theta}$ at $\boldsymbol{\theta}^{(\ell-1)}$.
    \ENDFOR\\[.3em]
    \OUTPUT  $\boldsymbol{\theta}^{(n_{\mathrm{iter}})}, \boldsymbol{\rho}^{(n_{\mathrm{iter}})}$.
    \end{algorithmic}
    \caption{UNHaP solver.}
    \label{algo}
\end{algorithm}

\looseness=-1
\noindent \textbf{Efficient parameter inference.} To allow UNHaP to scale to large physiological event detection applications, the estimation of the parameters $\boldsymbol{\theta}^{(\ell)}$ in the {\bf M}-step relies on the FaDIn framework \citep{staerman2023fadin}. This framework is adapted to capture delays between large events with general parametric kernels and efficient inference. It relies on three key ingredients: (1) the discretization of the timeline with a stepsize $\Delta$, (2) the use of finite support kernels $\phi_{ij}$ with length $W$ such that $\phi_{ij}(t)=0, \forall t \notin [0, W]$, and (3) precomputations terms for the $\ell_2$ loss, allowing to make the computational complexity of the optimization steps independent of the number of events.
Based on these ingredients, we add an index $\mathcal{G}$ to the losses, referring to the discretization grid on the previously introduced losses (\ref{eq:loss2}) and (\ref{eq:loss}).
For details on adapting this framework to our unmixing problem, we refer the reader to \autoref{subsec:discretization}.

\noindent \textbf{Minimization steps.} The {\bf E} and {\bf M} steps of \autoref{algo} are performed using gradient-based optimization on the losses $\mathcal{L}_{\mathcal{G}}(\boldsymbol{\theta}; \mathcal{Y}_T, \mathscr{F}_T )$ and $\bar{\mathcal{L}}_{\mathcal{G}}(\boldsymbol{\rho},\boldsymbol{\theta}; \mathscr{F}_T )$.
To improve the flexibility of the CEM procedure, we define a parameter $b$ that sets the number of optimization steps conducted on  $\boldsymbol{\theta}$ before updating $\boldsymbol{\rho}$. This parameter controls a trade-off between recovering the parameters of the two mixed processes and recovering the correct latent mixture structure.
The gradients w.r.t. each parameter are exhibited in the \autoref{subsec:gradients}.
The gradient of $\boldsymbol{\rho}$ requires the gradient of the precomputation terms w.r.t. $\boldsymbol{\rho}$. Therefore, these terms must be computed at each update of $\boldsymbol{\rho}$, \ie{} every $b$ optimization steps. The bottleneck of the computation cost of UNHaP is then the updates of precomputation terms. Given a number of iterations of our solver, say $n_{\mathrm{iter}}$, the total cost of the precomputation is dominated by
$O\big( \floor{n_{\mathrm{iter}}/ b} D^2L^2G \big)$, where $G$ is the number of elements of $\mathcal{G}$ and $L=\lfloor W/\Delta\rfloor$ is the number of elements of the grid used for the kernel discretization.

\textbf{Initialization with Moments Matching.} As it is generally the case when inferring Hawkes processes \citep{lemonnier2014nonparametric}, the loss $\mathcal{L}_{\mathcal{G}}$ is non-convex w.r.t. its parameters and may converge to a local minimum, thus yield sub-optimal parameters.
The quality of these minima strongly depends on the initialization scheme used for the initial value of the baselines and the kernel parameters.
A natural approach is to select them randomly.
However, this option can make the algorithm unstable and yield sub-optimal parameters as the solver can fall into irrelevant local minima.
Another option is to take advantage of the observed event distribution and perform moment matching to initialize the parameters.
We refer to this option as “Moments Matching initialization". Moment matching ensures that the moments of the observed distribution matches those of the parametric model at initialization. All the mathematical details and numerical experiments demonstrating the advantages of using Moments Matching are deferred in \autoref{subsec:mm} and \autoref{subsec:mm:appendix}, respectively.

\section{Numerical Validation}

In this section, we evaluate the benefits of UNHaP in recovering the structure of the mixture of latent variables and the parameters of the structured events on simulated data.
We also compare the performance of UNHaP with other PP solvers and show that it is more robust to noise while keeping a reasonable computational cost.

\subsection{Joint inference and unmixing with UNHaP}
\label{subsec:adv}

Based on simulated processes, we show that UNHaP jointly recovers the parameters of the Hawkes events and the mixture's latent variables in various noise settings.

\noindent \textbf{Simulation.} With the Immigration-Birth algorithm \citep{moller2005perfect,moller2006approximate}, we generate one-dimensional marked events in $[0, T] \times \mathcal{K}$ with $T=\{100, 1000, 10000 \}$ and $\mathcal K = [0, 1]$ from the mixture process with the following intensity function

{\small
\begin{equation}\label{eq:sim}
    \lambda(t, \kappa; \boldsymbol{\theta}) = \bigg (\mu + \alpha \sum_{t_n<t} Y_n\omega(\kappa_n) \phi(t-t_n; \eta) \bigg ) f^1(\kappa) + \tilde{\mu}~f^0(\kappa),
\end{equation}
}%
where $\omega(\kappa)=\kappa$ and $Y_n=1$ if $t_n$ is generated by the Hawkes process.
The intensity $\tilde{\mu}$ of the Poisson process is amenable to the noise level of the mixture process and $\alpha$ characterizes how strong the excitation structure is\footnote{the maximum authorized $\alpha$ parameter to have a stable process is such that $\alpha \mathbb{E}_{f^1}[\omega(\kappa)]= \frac{2 \alpha}{3}< 1$.}.
We denote $\boldsymbol{\alpha}=\alpha \mathbb{E}_{f^1}[\omega(\kappa)]$ the excitation level such that $\boldsymbol{\alpha} \to 1$ indicates a high excitation structure, with most events in the MMHP stemming from previous ones, while $\boldsymbol{\alpha} \to 0$ indicates no structure, as the process is almost a Poisson process.
$f^0$ and $f^1$ are the marks' distributions and we set $f^1(\kappa) = 2\kappa ~$ to account for a linear mark distribution for structured events.
For the noisy marks, we consider two settings: one linear with $f^0(\kappa)=2(1-x)$ and one uniform with $f^0(\kappa) = 1$.
These two cases correspond to different information levels present in the marks on the probability of being a true event.
The excitation kernel $\phi (\cdot; \eta)$ is chosen as a truncated Gaussian kernel, to model delays between the events.
With $\eta=(m,\sigma)$, it reads
\begin{equation*}
    \phi(\cdot; \eta)  = \frac{1}{\sigma} \frac{\gamma\left(\frac{\cdot-m}{\sigma}\right)}{F\left(\frac{W-m}{\sigma}\right)-F\left(\frac{-m}{\sigma}\right)} \1[0\leq \cdot \leq W],
\end{equation*}

where $W$ is the kernel length and $\gamma$ (resp. $F$) is the probability density function (resp. cumulative distribution function) of the standard normal distribution.
In our experiments, we set $\eta=(0.5, 0.1)$.

\begin{figure}[!t]
    \centering
    \begin{tabular}{c m{2em} c}
    \multicolumn{3}{c}{\includegraphics[trim=0cm 0cm 0cm 0cm, scale=0.55]{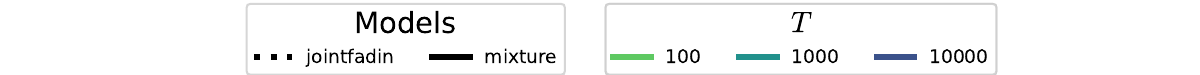} \vspace{-0.1cm}}
           \\
        \includegraphics[trim=0cm 0cm 0cm 0cm,scale=0.4]{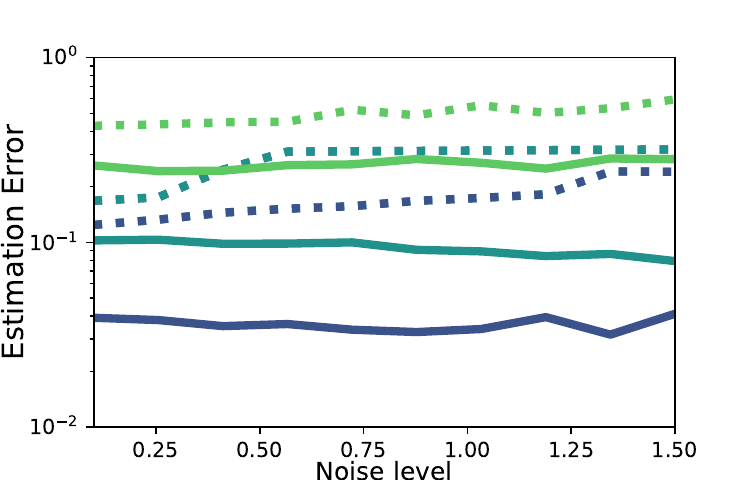} &&
        \includegraphics[trim=0cm 0cm 0cm 0cm,scale=0.4]{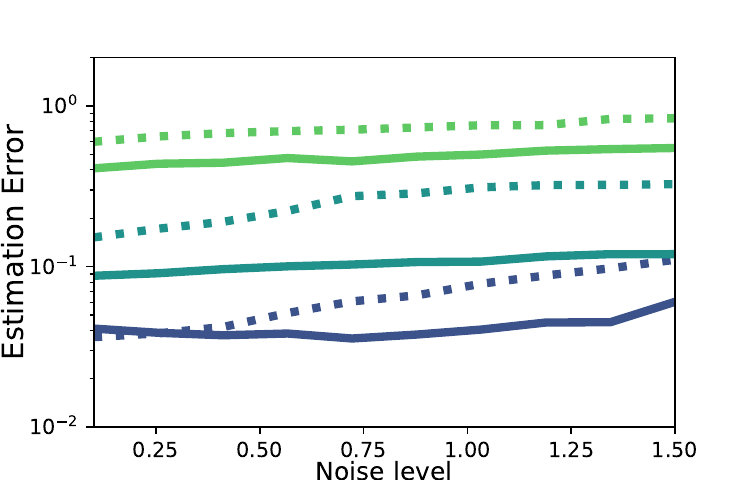}
    \end{tabular}
    \caption{Parameters estimation errors for UNHaP and ``jointfadin'' for varying $T$ w.r.t. different values of $\tilde{\mu}$ with linear (\emph{left}) and uniform (\emph{right}) distributions on noisy marks.}
    \label{fig:infer}
    \vspace{-0.5cm}
\end{figure}

\noindent \textbf{Robust parameter inference with UNHaP.}
To highlight the robustness of the Hawkes excitation structure recovery with UNHaP and the necessity to infer the mixture parameters, we compare the parameter recovery for different noise levels $\tilde{\mu} \in [0.1, 1.5]$.
We set $\mu=0.8$, $\alpha=1.45$, which correspond to a process with clear structure.

We infer the MMHP's parameters $\boldsymbol{\theta} = \{\mu, \alpha, m, \sigma\}$ with UNHaP and compare our results with a marked version of FaDIn, which we called ``JointFaDIn''.
We set $\Delta=0.01$ and $W=1$ with $10000$ optimization steps for UNHaP and JointFaDIn.
The number of iterations chosen between two updates of $\hat{\rho}$ is set to $b=200$ according to the sensitivity study depicted in \autoref{subsec:batch}.
\autoref{fig:infer} reports the median value over 100 repetitions of $||\hat{\boldsymbol{\theta}} - \boldsymbol{\theta} ||_2$, reflecting the error between the estimates and their actual values, for the linear marks (left) and uniform marks (right) settings.
UNHaP outperforms JointFaDIn in all settings while being more robust w.r.t. the noise level $\tilde{\mu}$, as the performances remain constant.
This experiment shows that accounting for the mixture's latent variables is crucial to recovering the parameters of the structured events.
We also see that the linear mark distribution allows for better parameter recovery, as it is more informative to infer the mixture's latent variables.

\noindent \textbf{UNHaP recovers the mixture structure.} To show the performance of UNHaP to classify the observed events between the spurious and structured ones, we use the simulated processes defined above, varying $\alpha \in [0, 1]$.
Here, we set $\mu=0.4$ and $\tilde{\mu}=0.1$.
Experiments varying the noise levels ($\tilde{\mu}=0.5$ and $\tilde{\mu}=1$) are presented in \autoref{fig:rho:appendix}.

\begin{figure}[!t]
    \centering
    \begin{tabular}{ccc}
    \includegraphics[trim=0cm 0cm 0cm 0cm,scale=0.45]{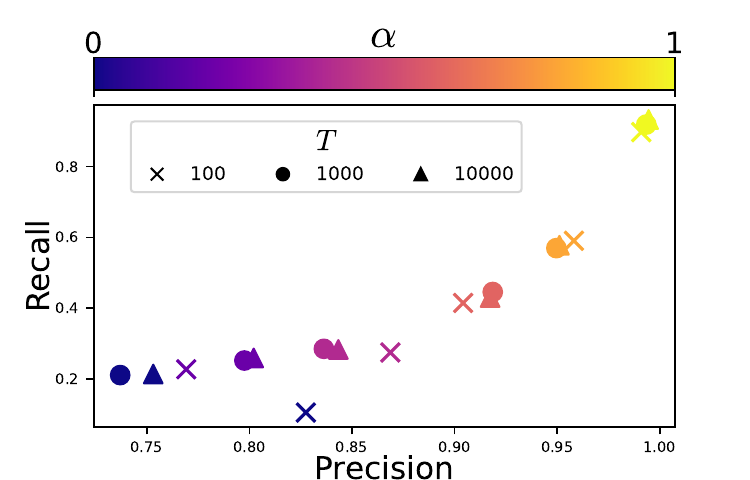} &
    \includegraphics[trim=0cm 0cm 0cm 0.5cm,scale=0.45]{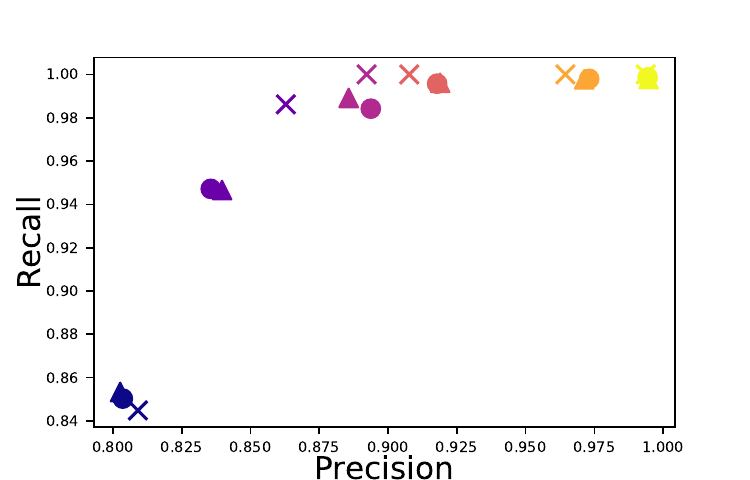}
    \end{tabular}
    \caption{Precision/Recall values for the estimation of $\rho$ for different values of $T$ w.r.t. $\alpha$ with linear (\emph{left}) and uniform (\emph{right}) distributions on noisy marks.}
    \label{fig:rho}
\end{figure}

We consider the mixture parameter $\boldsymbol{\rho}$ inferred with UNHaP in the case of structured  (linear setting) and unstructured (uniform) noise marks.
We set $\Delta=0.01$, $W=1$ and $b=200$ with $10000$ optimization steps for UNHaP.
In \autoref{fig:rho}, we report the Precision and Recall scores of the estimated mixture parameter $\hat{\rho}$ w.r.t. the ground truth.
We can see the convergence $\hat{\rho}$ towards the true $\rho^*$ when the excitation structure grows in both cases.
When $\alpha$ is small and the excitation structure absent, only the mark distribution may distinguish between the events stemming from $\mu$ and $\tilde{\mu}$.
\autoref{fig:rho} shows that the accuracy of $\hat{\rho}$ stays high for a small $\alpha$ when mark densities are different (right), but it is challenging when they overlap (left).

\subsection{Benchmarking inference and computation time}
\label{sect:bench}

We compare UNHaP with various Hawkes process solvers by assessing approaches' statistical and computational efficiency in the case of simulated noisy and non-noisy data. Considering the few models and open-sourced code available in the marked Hawkes process area, we compare UNHaP with popular unmarked Hawkes solvers.
The parametric approaches we compare with are 1) FaDIn \citep{staerman2023fadin}; 2) the Neural Hawkes Process (Neural Hawkes; \citealp{mei2017neural}) where authors model the intensity function with an LSTM module; and 3) Tripp \citep{NEURIPS2020_00ac8ed3}, where a triangular map is used to approximate the compensator function, i.e., the integral of the intensity function.

{\renewcommand{\arraystretch}{1.5} 
{\setlength{\tabcolsep}{0.3cm}
\begin{table*}[!ht]
\begin{center}
{\scriptsize
\begin{tabular}{c|ccc || ccc}
& \multicolumn{3}{c}{NLL} &  \multicolumn{3}{c}{Computation time (s)} \\
\hline
$T$ & 100 & 500 & 1000 & 100 & 500 & 1000\\
\hline
  UNHaP  &  \textbf{0.624 $\pm$ 0.31}& \textbf{0.447 $\pm$ 0.12}  &\textbf{0.346 $\pm$ 0.03} & 96.2 $\pm$ 4.5 & 109.6 $\pm$ 5.9 & 117.4 $\pm$ 5.8 \\
  FaDIn & 2.445 $\pm$ 0.19 & 2.442 $\pm$ 0.1  & 2.441 $\pm$ 0.14 & 41.3 $\pm$ 19.4 & 32.5 $\pm$ 12.8 & 30.9 $\pm$ 5.9 \\
  Tripp & 4.27 $\pm$ 0.62 & 2.137 $\pm$ 0.18 & 1.555 $\pm$ 0.07 & 44.6 $\pm$ 6.7 & 50.9 $\pm$ 3.7 & 55.3 $\pm$ 3.5 \\
  Neural Hawkes & 2.006 $\pm$ 0.7 & 1.574 $\pm$ 0.45 & 1.141 $\pm$ 0.2 & 43.4 $\pm$ 16.8 & 171.8 $\pm$ 38.1 & 183.3 $\pm$ 30.7 \\
\hline
\end{tabular}
}
\end{center}
\caption{Mean $\pm$ standard deviation (over ten runs) of the Negative Log-Likelihood (NLL) \textbf{on marked events in noisy settings} for various models and various sizes of events sequence.}
\label{tab:bench}
\end{table*}}}

We simulate a marked Hawkes process in a high noise setting, where noisy events have a small mark compared to the Hawkes events.
Its intensity function is defined as in (\ref{eq:sim}), with a linear mark distribution in [0, 1].
We also simulate a Poisson Process for the noisy events, with a uniform mark distribution in [0, 0.2].
Therefore, $\omega(\kappa)=\kappa$, $f^1(\kappa) = 2\kappa$ and $f^0(\kappa) = \1[0\leq \kappa \leq 0.2]$.
We set $\mu=0.1$, $\alpha=1$, imposing a high excitation phenomenon, and $\tilde{\mu}=1$, corresponding to a high-noise setting.

We then conduct inference on the intensity function of the underlying Hawkes processes using UNHaP and the three aforementioned methods, $\Delta=0.01$ applied consistently across all discrete approaches and $W=1$ for FaDIn and UNHaP.
This experimental procedure is replicated for various values of $T \in \{10, 500, 1000\}$.
The NLL is computed on a test set simulated with parameters identical to the training data.
The median NLL over ten runs and the computation time are displayed in \autoref{tab:bench}.

In this marked and noisy setting, UNHaP demonstrates a statistical superiority over all methods.
This outcome aligns with expectations in a parametric approach when the utilized kernel belongs to the same family as the one used for event simulation.
It is essential to highlight that these results stem from analyzing a single (long) data sequence, contributing to the subpar statistical performance of Neural Hawkes, which excels in scenarios involving numerous repetitions of short sequences due to the considerable number of parameters requiring inference.
From a computational time standpoint, UNHaP takes much longer than the other methods but is the only one converging to an accurate result.
It is slower compared to FaDIn, which is expected due to the alternate minimization scheme, which performs repeated parameter inference using a procedure similar to FaDIn.
UNHaP is the only successful solver in the noisy data context at a reasonable computation cost.
In an unmarked setting, UNHaP performs on par with the other methods, with a slight advantage in noisy settings; see \autoref{tab:bench_nomark}.



\vspace{-0.2cm}
\section{Application to Physiological Data}\label{sec:ecg}
\vspace{-0.2cm}
To demonstrate the usefulness of UNHaP in real-world applications, we use it to characterize the inter-event interval distribution in ECG and gait data. Statistics derived from ECG inter-beat intervals, such as the heart rate (HR) and the heart rate variability (HRV), are central in diagnosing heart-related health issues, like arrhythmia or atrial fibrillation~\citep{shaffer2017overview}.
Similarly, the study of a person's gait with inertial measurement units (IMU) is essential in diagnosing pathologies like Parkinson's disease or strokes~\citep{truong2019data}, in particular by analyzing the inter-step time intervals.
Computing these statistics requires a robust detection of heartbeats~\citep{berkaya2018survey} or steps~\citep{oudre2018template}.
Classical domain-specific methods are typically used~\citep{Pan1985, elgendi2013fast, hamilton2002open}, in combination with heavily tailored post-processing steps~\citep{merdjanovska2022comprehensive,oudre2018template} to cope with spurious event detection resulting from noisy signals.
The design of such methods is cumbersome, requires domain expertise, and does not generalize well.

A more automatized approach to detect events is to use Convolutional Dictionary Learning (CDL;~\citealp{latour2018multivariate}).
While it is more domain-agnostic than classical methods, this method is even more prone to spurious event detection.
UNHaP circumvents this issue by post-processing the detected events to separate structured events from spurious ones.
In the following, we use UNHaP to post-process ECG and gait events detected using CDL.
We show on ECG data from the \textit{vitaldb}~\citep{lee2022vitaldb} and gait data from \cite{truong2019data} that our generic methodology reaches performance on par with state-of-the-art, heavily tailored methods.
Our results showcase that UNHaP filters out noisy events and that the inferred parameters are coherent with the physiological data.

\begin{figure*}[t]
    \centering
    \includegraphics[width=0.9\textwidth]{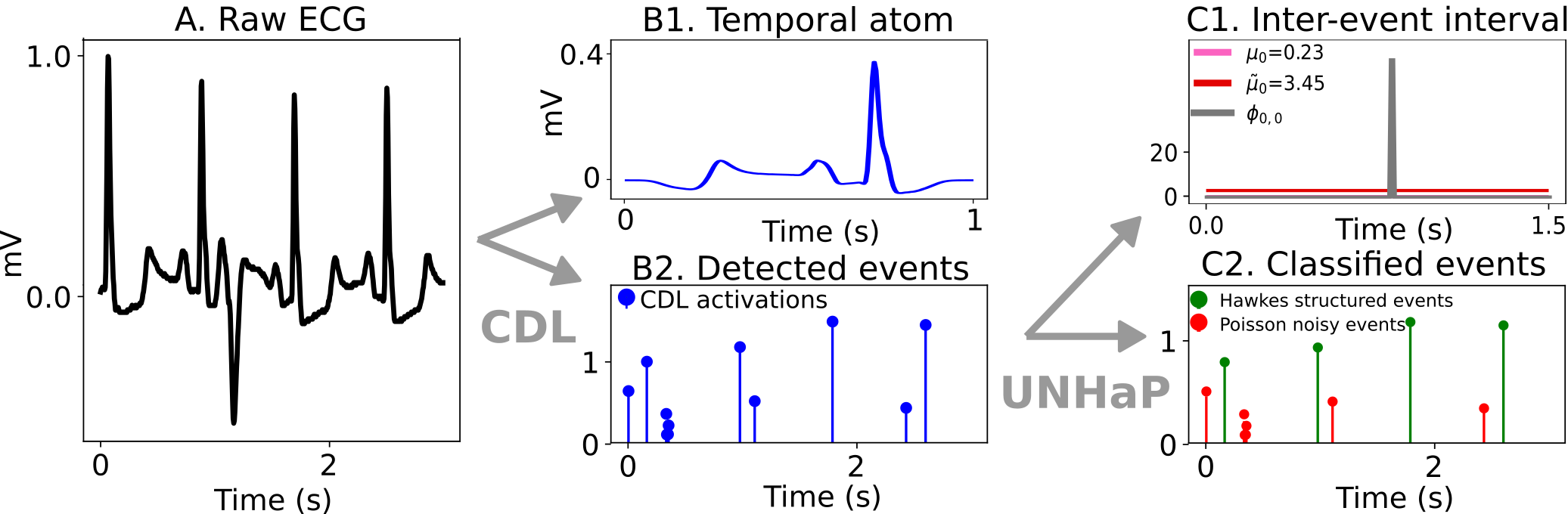}
    \caption{%
        \looseness=-1%
        Experimental pipeline on ECG Data. \textbf{(A)} Sub-sample of raw ECG plot. \textbf{(B)} Output of Convolutional Dictionary Learning algorithm: \textbf{(B1)} learned temporal atom representing one heartbeat, \textbf{(B2)} detected events on the time interval. \textbf{(C)} Output of UNHaP: \textbf{(C1)} Estimated Hawkes parameters: noise baseline (red), baseline (pink), and kernel (grey). The kernel is very close to the ground truth (orange dashed). \textbf{(C2)} Unmixing output $\rho$: events were classified either belonging to the Hawkes Process (\textcolor{structured}{\bf green}) or as spurious noisy events (\textcolor{spurious}{\bf red}).}
    \label{fig:ecg}
\end{figure*}

\noindent \textbf{Experimental pipeline for CDL+UNHaP method.}
We used the same method for ECG and gait recordings. In what follows, it is illustrated on ECG recordings. The proposed method relies on the CDL algorithm from the Python library \href{https://alphacsc.github.io/}{\texttt{alphacsc}} \citep{latour2018multivariate} to detect events.
Denote by $X$ an ECG slot, CDL decomposes it as a convolution between a dictionary of temporal atoms $D$ and a sparse temporal activation vector $Z$: $X = Z*D + \varepsilon$.
Starting from the ECG signal depicted in \autoref{fig:ecg} (A), panel (B1) shows the learned temporal atom on ECG, and panel (B2) shows the activation vector $Z$ from ECG window obtained with CDL.
There are non-zero activations for each beat, but they are mixed with noisy activations in $Z$.
Handcrafted thresholding methods would typically be used here to remove noisy activations from $Z$, but would need to be adapted for each recording \citep{Allain2022,staerman2023fadin}.
Instead, we process the raw activation vector $Z$ with the proposed UNHaP method.
The solver separates the heartbeat Hawkes Process from the noisy activations (\autoref{fig:ecg} (C2)) and estimates the inter-burst interval (\autoref{fig:ecg} (C1)).
The mean (respectively the standard deviation) of the parameterized truncated Gaussian $\phi$ estimates the mean inter-beat interval (respectively the heart rate variability) on the ECG slot $X$.
With this example, we see that UNHaP successfully detects the structured events from the noisy ones, providing a good estimate of the inter-beat distribution. Additional experimental details are provided in \autoref{sect:supp_physio}.


\noindent \textbf{Results.}
We compare several estimators of the inter-beat and inter-step interval distributions, including the developed pipeline UNHaP and with FaDIn \citep{staerman2023fadin} to post-process the detected events.
We also compare several domain specific libraries: \href{https://github.com/PGomes92/pyhrv/tree/master}{\texttt{pyHRV}} \citep{gomes_pgomes92pyhrv_2024}, \href{https://neuropsychology.github.io/NeuroKit/introduction.html#}{\texttt{Neurokit}} \citep{makowski_neurokit2_2021}, which are tailored to ECG, and Template Matching \citep{oudre2018template}, a method specifically tailored for gait detection.

For ECG, the mean inter-beat interval obtained with these estimators is compared to the ground truth, which is given in the dataset.
We average each estimator's absolute and relative absolute errors over the 19 ECG recordings, and report the results in \autoref{tab:physio}.
UNHaP, pyHRV, and Neurokit have equivalent performance, all providing good heart rate estimates, even though our method has a increased variance.
Another interesting finding is that while working with the same detected events, UNHaP vastly outperforms FaDIn.
This highlights again the benefit of our mixture model to separate structured events --here quasi-periodic-- from spurious ones.

We applied the same pipelines to gait recordings.
Results in \autoref{tab:physio} show that UNHaP performs comparably to the  state-of-the-art on gait data. Methods designed for ECG (pyHRV and Neurokit) fail to provide accurate estimates of the inter-step interval.



{\renewcommand{\arraystretch}{1.2} 
\begin{table}[!ht]
\centering
{\setlength{\tabcolsep}{0.18cm}
\begin{tabular}{c|c|c|c|c|c|c}
    \multicolumn{2}{c|}{} & \makecell{\bf \footnotesize CDL\\\footnotesize + \bf UNHaP} & \makecell{\footnotesize CDL\\\footnotesize + FaDIn} & { \footnotesize pyHRV} & {\footnotesize Neurokit} & \makecell{\footnotesize Template\\ \footnotesize Matching}\\
\hline
   \multirow{2}{*}{ECG}& {\small AE (beats/min)} & 0.61$\pm$0.93  & 15.44$\pm$19.39 & 0.57$\pm$0.14 & 0.45$\pm$0.07 & \multirow{2}{*}{NA}\\
     & {\small RAE}  & 0.009$\pm$0.012 & 0.192$\pm$0.227 & 0.008$\pm$0.002 & 0.006$\pm$0.001\\\hline\\[-1em]\hline
    Gait &
    {\small MAE (seconds)}
    & 0.04 & 1.2 & 0.67 & 0.57  & 0.07\\
\end{tabular}}
\caption{Absolute Error (AE) and Relative Absolute Error (RAE) of the estimated average inter-beat interval (mean $\pm$ std) on the \textit{vitaldb} dataset (\emph{top}). Median Absolute Error (MAE) of the estimated average inter-step interval on gait (\emph{bottom}).}
\label{tab:physio}
\vspace{-0.8cm}
\end{table}}

\section{Discussion}

Having defined the challenge of distinguishing unstructured Poisson processes from structured events, this work introduces UNHaP -- a model built upon a mixture marked Hawkes process designed to disentangle noise from structured events. UNHaP utilizes latent variables to represent the mixture of the two marked processes to eliminate spurious events. This is achieved by minimizing an ERM-inspired least squares loss, incorporating finite-support kernels and discretization, and ensuring reasonable computational costs. Additionally, UNHaP accommodates using any parametric form of triggering kernels, making it particularly pertinent for monitoring ECG heart rate. We demonstrate the benefits of using our unmixing models rather than the traditional Hawkes process models with simulated and real-world ECG and gait data. This illustrates UNHaP's universality compared to existing methods. The pipeline developed here is unsupervised and requires no pre-processing or data adjustment. It is agnostic and meant to be robust on a wide range of data modalities.


\section*{Acknowledgments}

The authors thank Jérôme Cartailler et Jade Perdereau for proofreading the paper. Virginie Loison was supported by the ANR EBUL (ANR-23-CE23-0001) and Guillaume Staerman by the chair BrAIN (ANR-20-CHIA-0016).

\bibliographystyle{apalike}
\bibliography{biblio}

\newpage
\appendix

\counterwithin{figure}{section}
\def\thefigure{\thesection.\arabic{figure}}
\counterwithin{table}{section}
\def\thetable{\thesection.\arabic{table}}

\section{Technical Details}

\subsection{Detailing UNHaP loss with discretization}\label{subsec:discretization}

In the following, we assume that the functions $\omega_{ij}(\cdot)$ are identical for $1\leq i,j\leq D$ and denote it by $\omega(\cdot)$.


\noindent \textbf{Discretization and finite support kernels.} Motivated by computational efficiency and the use of general parametric kernels, we adopt a setting similar to the one recently proposed by~\cite{staerman2023fadin}.
First, we discretize the time by projecting each event time $t^i_n$ on a regular grid $\mathcal{G}=\{0, \Delta, 2\Delta, \ldots, G\Delta \}$, where $G = \floor{\frac T \Delta}$. We refer to $\Delta$ as the stepsize of the discretization and denote by $\widetilde{\mathscr{F}}_T^i$ the set of projected events of $\mathscr{F}_T^i$ on the grid $\mathcal{G}$.
Second, we suppose the length of the kernels $\phi_{ij}$ to be finite. This assumption is consistent with scenarios in which an event's impact is limited to a relatively short time frame in the future. Examples of such applications include neuroscience \citep{Allain2022} or high-frequency trading \citep{bacry2015hawkes}. We denote by $W$ the length of the kernel's support kernel, such that $\forall i,j, \; \forall t \notin [0, W], \phi_{ij}(t) = 0$. The size of the kernel of the discrete grid is then equal to $L = \floor{\frac W \Delta}$.
With these two key features, the intensity  boils down to

\begin{align*}
    \bar\lambda_i^1([s], \kappa;\boldsymbol{\theta}_1) = \bigg ( \mu_i + \sum_{j=1}^{D}\sum_{\tau=1}^{L} \phi_{ij}^{\Delta}[\tau] \tilde{z_j}[s-\tau]  \bigg) f_i^1(\kappa),
\end{align*}

where $s\in \intervalleEntier{0}{G} $ and $\phi^{\Delta}_{ij}[\cdot], \tilde{z}_j[\cdot]$ are vector notations. Precisely, $\phi_{ij}^{\Delta}[s] = \phi_{ij}(s\Delta)$ and $\tilde{z}_{j}[s] = \sum_{t^j_n} \rho^j_n ~ \omega(\kappa^j_n) ~\1[\braces{|t_n^j - s\Delta|\leq \frac{\Delta}{2}}]$. For notation convenience, we introduce the vectors $\rho^j[\cdot], z_j[\cdot]$ such that $\rho^j[s] = z_j[s] =0$ when there is no events at location $s$ and to $\rho^j[s]=\rho^j_n,z_j[s]=\omega(\kappa_n^j) $ if there is an event $t_n^j$ at position $s$. Therefore, $\tilde{z_j}$ can be written as $\tilde{z_j} = \rho^j \odot z_j \in \mathbb{R}_+^{G+1}$ where $\odot$ is the Hadamard product. The computation of the intensity function is more efficient in the discrete approach, leveraging discrete convolutions with a worst-case complexity that scales as $O(N_{g}(T) L)$, where $N_{g}(T)=\sum_{i=1}^{D} N_{g_i}(T)$ is the total number of events, contrasting with the quadratic complexity w.r.t. $N_g(T)$ in general parametric kernels. The bias introduced by the discretization setting is negligible in most cases \citep{Kirchner2016,kirchner2018nonparametric,staerman2023fadin}.

\looseness=-1
\noindent \textbf{Efficient Inference.} Our approach aims at minimizing the discretized version of $\bar{\mathcal{L}}(\boldsymbol{\rho}; \boldsymbol{\theta}, \mathscr F_T)$  and $\mathcal{L}(\boldsymbol{\theta}; \mathcal{Y}_T, \mathscr{F}_T)$ according to the latent mixture' parameters $\boldsymbol{\rho}$ and the process's parameters  $\boldsymbol{\theta}$. Given the previous notations, we get

{\small
\begin{align*}
\bar{\mathcal{L}}^i_{\mathcal{G}}(\boldsymbol{\rho}, \boldsymbol{\theta}, \widetilde{\mathscr F}_T) &=
T(H_i^1\mu_i^2  +H_i^0\tilde{\mu}_i^2  )
+ 2\Delta H_i^1\mu_i   \sum_{j=1}^{D} \sum_{\tau=1}^{L} \phi_{ij}^{\Delta}[\tau] \widetilde{\Phi}_j(\tau; G) \\&
+ \Delta H_i^1\sum_{j,k} \sum_{\tau=1}^{L}\sum_{\tau'=1}^{L} \phi_{ij}^{\Delta}[\tau] \phi_{ik}^{\Delta}[\tau'] \widetilde{\Psi}_{j, k}(\tau, \tau'; G)
+ \Delta \sum_{j=1}^{D} \sum_{\tau=1}^{L} \phi_{ij}^{\Delta}[\tau]^2 ~\widetilde{\Xi}_j(\tau; G) \\&
- 2 \Bigg (\tilde{\mu}_i\sum_{(\tilde{t}^i_n, \kappa_n^i) \in \widetilde{\mathscr F}_T^i} f_i^0(\kappa^i_n) \bigg (1-\rho^i\bracks{\frac{\tilde{t}_n^i}{\Delta}} \bigg )  \\&
 +    \mu_i \sum_{(\tilde{t}^i_n, \kappa_n^i) \in \widetilde{\mathscr F}_T^i}  f_i^1(\kappa^i_n)\rho^i\bracks{\frac{\tilde{t}_n^i}{\Delta}}
 + \sum_{j=1}^{D} \sum_{\tau=1}^{L}\phi^{\Delta}_{ij}[\tau] \widetilde{\Phi}_{j} (\tau; \widetilde{\mathscr{F}}_T^i)
\Bigg ),
\end{align*}
}
where $H_i^\ell = \int_{\mathcal{K}} (f^\ell_i(\kappa))^2 ~\mathrm{d} \kappa$ for $\ell\in \{0, 1 \}$ and $\; \widetilde{\Phi}_{j}(\tau;  G) = \sum_{s=1}^{G} \tilde{z}_j[s-\tau] $, $ \widetilde{\Psi}_{jk}(\tau, \tau'; G) =\sum_{s=1}^{G} \tilde{z}_j[s-\tau] \tilde{z}_k[s-\tau']$,
$\; \widetilde{\Xi}_j(\tau; G)=\sum_{s=1}^{G} \big( z_j^2[s-\tau] \rho^j[s-\tau] -\tilde{z}_j^2[s-\tau] \big)$ and $\widetilde{\Phi}_j(\tau; \widetilde{\mathscr{F}}_T^i) =\sum_{(\tilde{t}_n^i, \kappa_n^i) \in \widetilde{\mathscr{F}}_T^i} f_i^1(\kappa^i_n) \rho^{i}\bracks{\frac{\tilde{t}_n^i}{\Delta}} \tilde{z}_j\bracks{\frac{\tilde{t}_n^i}{\Delta}-\tau}$. Conditionally to the knowledge of $\boldsymbol{\rho}$, these last four terms can be precomputed, removing the computational complexity's dependency on the number of events (here represented by the grid) during the optimization on parameters $\boldsymbol{\theta}$. The cost of computing $\widetilde{\Psi}_{j,k}(\cdot, \cdot; G)$ is dominating and requires $O(G)$ operations for each $(\tau,\tau')$ and $(j,k)$ leading to $O(D^2L^2G)$ as in the FaDIn framework. Note that  the loss $\mathcal{L}_{\mathcal{G}}(\boldsymbol{\theta}; \mathscr{F}_T, \mathcal{Y}_T)$  can be derived identically, one may just replace the $\rho_n^i$ by $Y_n^i=\mathbb{I}\{\rho_n^i > 1/2 \}$ and removing the fourth term.
\subsection{Initialization with Moments Matching} \label{subsec:mm}

Moment matching ensures that the moment of the observed distribution matches the moment of the parametric model with the initial parameter.
Let us consider a multivariate marked Hawkes Process of ground intensity functions $\{\lambda_{g_i}\}$ and ground counting processes $N_{g_1}, \ldots, N_{g_D}$ being equal to the number of observed events on time interval $[0, T]$.
The proposed initialization method relies on choosing initial parameters such that the empirical process expectation  is equal to the expectation of the model, \emph{i.e.}
\begin{equation}\label{eq:moment_method}
    N_{g_i}(T) = \mathbb{E}[N_{g_i}(T)]
    = \int_0^T\lambda_{g_i}(t)~\mathrm{d}t.
\end{equation}
This system is not fully determined as we only have one equation for multiple unknown variables.
To compute a simple solution for this system, we make some extra assumptions.
First, we consider that all $\rho_n^i$ are equal to $\frac12$.
With this, we get $N^0_{g_i}(T) = \frac{N_{g_i}(T)}2$ and thus we can compute a moment matching value  $\tilde{\mu}_i^{\text{m}}$ since

\begin{equation*}
\frac{N_{g_i}(T)}{2} = \int_0^T {\lambda}_{g_i}^0(s) \mathrm{d}s = T {\tilde{\mu}}_i \Rightarrow \tilde{\mu}_i^{\text{m}} = \frac{N_{g_i}(T)}{2T}.
\end{equation*}

Similarly, we get $N_i^1(T) = \frac{N_{g_i}(T)}2$ and thus, as $N_i^1(T) = \int_0^T\lambda_{g_i}^0(s)\mathrm{d}s$, we get
\begin{equation*}
\frac{N_{g_i}(T)}{2} = \mu_i T + \sum_{j=1}^D\alpha_{i, j}^{\text{m}}\sum_{(\tilde{t}_n^j, \kappa^j_n) \in \widetilde{\mathscr{F}}_T^j} \omega(\kappa_n^j).
\end{equation*}
Once again, we have only one equation with $D+1$ unknown parameters. We choose to assume that each parameter will generate the same amount of events, leading to
\begin{equation*}
\mu_i^{\text{m}} = \frac{N_{g_i}(T)}{2T(D+1)},
\end{equation*}
and
\begin{equation*}
\alpha_{i, j}^{\text{m}} = \frac{N_{g_i}(T)}{2T(D+1)\sum_{(\tilde{t}_n^j, \kappa^j_n) \in \widetilde{\mathscr{F}}_T^j} \omega(\kappa_n^j)}.
\end{equation*}

Replacing these values for $\tilde{\mu}_i^m, {\mu}_i^m$, and $\alpha_{i, j}^m$ into (\ref{eq:moment_method}) ensures that the number of events' expectation for the parametric model matches the one from the observed process.
The other kernel parameters are initialized using the method of moments on the delay between events. Denoting by $\delta t^{i, j}_n$ the delay between $t^{i}_n$ and the time of occurrence of the last event in channel $j$ before $t^i_n$

\begin{equation}\label{eq:mm_deltatmax}
\delta t^{i, j}_n = t_n^i - \max\{t | t \in \mathscr{F}^j_T, W < t < t_n^i\}. \\
\end{equation}

For the truncated Gaussian kernel, defined in \autoref{subsec:adv}, the initial mean $m_{i,j}^{\text{m}}$ and standard deviation $\sigma_{i,j}^{\text{m}}$ are
\begin{align*}
m^{\text{m}}_{i, j} &= \frac{1}{N_{g_i}(T)}\sum_{t_n^i \in \mathscr{F}_T^i} \delta t^{i, j}_n, \\
\sigma^{\text{m}}_{i, j} &=\sqrt{\dfrac{\sum_{t_n^i \in \mathscr{F}_T^i} (\delta t^{i, j}_n - m^{\text{m}}_{i, j})^2}{N_{g_i}(T) - 1}}.
\end{align*}
For the raised cosine kernel, detailed in the \autoref{subsec:mm:appendix}, initial parameters $u_{i, j}^{\text{m}}$ and $s_{i,j}^{\text{m}}$ are computed similarly
\begin{align*}
u^{\text{m}}_{i, j} &= \max{(0, m^{\text{m}}_{i, j} - \sigma^{\text{m}}_{i, j})}, \\
s^{\text{m}}_{i, j} &= \sigma_{i, j}^{\text{m}}.
\end{align*}

\begin{figure*}[!ht]
    \centering
    \begin{tabular}{cc}
         \includegraphics[scale=0.5]{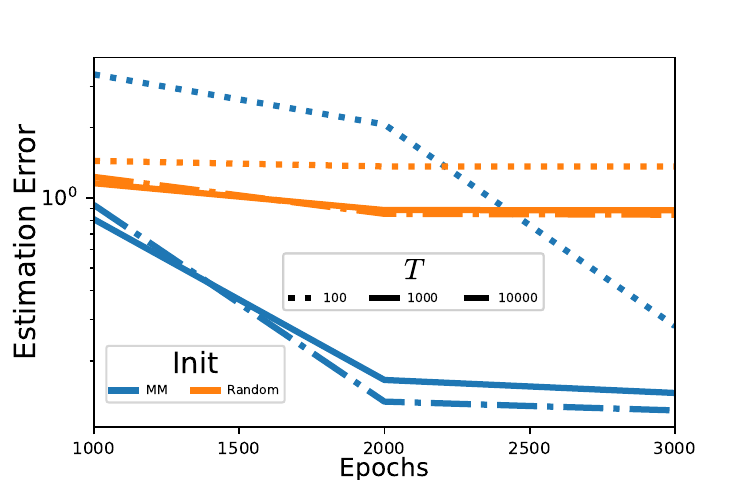} & \includegraphics[scale=0.5]{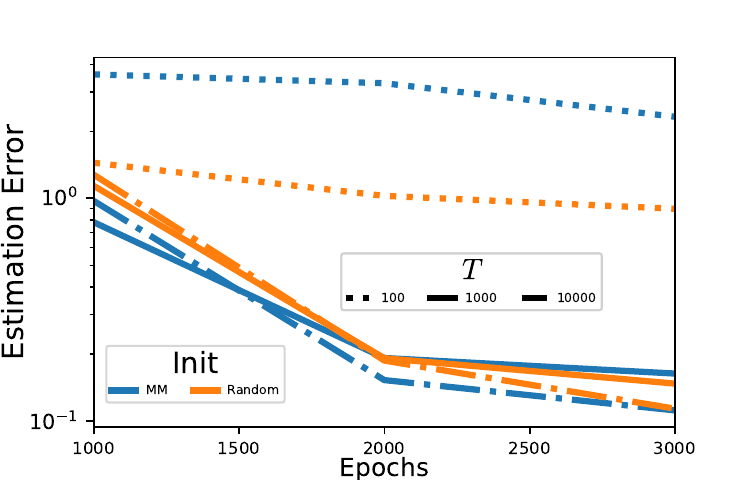}
    \end{tabular}
    \caption{Hawkes parameters estimation error using UNHaP with a raised cosine (left) and a truncated Gaussian (right) kernels, Moments Matching (blue), and Random (orange) initializations for varying size sequences.}
    \label{fig:moment_matching:appendix}
\end{figure*}

The benefits of this approach is supported by the numerical studies in \autoref{subsec:mm:appendix}. The moment matching initialization significantly improves convergences and lowers the risk of converging to irrelevant parameter values in the case of the raised cosine, while it behaves comparably in the case of the truncated Gaussian, see \autoref{fig:moment_matching:appendix}.

For very noisy settings, where noisy events are very close to Hawkes process events in time, using the $\delta t_n^{i,j}$ defined in (\ref{eq:mm_deltatmax}) leads to poor performance of UNHaP. This is because $\delta t_n^{i,j}$ is then tiny, leading to a very small initial mean, from which the solver has trouble converging to correct values. We circumvented this issue by computing $\delta t_n^{i,j}$ with a mean instead of a maximum.

\begin{equation}\label{eq:mm_deltatmean}
\delta t^{i, j}_n = t_n^i - \frac{1}{\#\{t \in \mathscr{F}^j_T, W < t < t_n^i\}} \sum_{t \in \mathscr{F}^j_T, W < t < t_n^i} t. \\
\end{equation}

\subsection{Gradients of the UNHaP loss}\label{subsec:gradients}

This part present the derivation of the gradients of the loss function minimized by UNHaP for each parameter.

\textbf{Gradient of the baseline.} For any $m\in \{1, \ldots, D \}$, we get

\begin{align*}
    \frac{\partial\bar{\mathcal{L}}_{\mathcal{G}}}{\partial \mu_{m}}&=
       2 TH_m^1 \mu_m    + 2\Delta  H^1_m\sum_{j=1}^{D} \sum_{\tau=1}^{L} \phi_{mj}^\Delta[\tau]\widetilde{\Phi}_{j}(\tau;  G)
      - 2 \sum_{(\tilde{t}^m_n, \kappa_n^m) \in \mathscr F_T^m} f^1_m(\kappa_n^m) \rho^m\bracks{\frac{\tilde{t}_n^m}{\Delta}}
\end{align*}

\textbf{Gradient of the noise baseline.}  For any $m\in \{1, \ldots, D \}$, we get

\begin{align*}
   \frac{\partial\bar{\mathcal{L}}_{\mathcal{G}}}{\partial \tilde{\mu}_{m}} &=
       2 T H_m^0 \tilde{\mu}_m
       -  2 \sum_{(\tilde{t}^m_n, \kappa_n^m) \in \mathscr F_T^m} f^0_m(\kappa_n^m) \bigg ( 1-\rho^m\bracks{\frac{\tilde{t}_n^m}{\Delta}} \bigg ).
\end{align*}

\textbf{Gradient of the excitation kernel parameters.} For any tuple $(m,l)\in \{1,\ldots, D\}^2$, the gradient of  $\eta_{ml}$ is
{\small 
\begin{align*}
        \frac{\partial\bar{\mathcal{L}}_{\mathcal{G}}}{\partial \eta_{ml}}= &  2\Delta H_m^1 \mu_m   \sum_{\tau=1}^{L} \frac{\partial \phi_{ml}^\Delta[\tau]}{\partial \eta_{ml}}~ \widetilde{\Phi}_l(\tau; G)
        + 2\Delta H_m^1 \sum_{k=1}^{D} \sum_{\tau=1}^{L} \sum_{\tau'=1}^{L} \phi_{mk}^\Delta[\tau'] \frac{\partial \phi_{ml}^\Delta[\tau]}{\partial \eta_{ml}}~\widetilde{\Psi}_{l,k}(\tau, \tau'; G) \\& \\&
        + 2\Delta \sum_{\tau=1}^{L} \frac{\partial \phi_{ml}^\Delta[\tau]}{\partial\eta_{ml}} \phi_{ml}^\Delta[\tau] ~\widetilde{\Xi}_l(\tau; G)  -2 \sum_{\tau=1}^{L} \frac{\partial \phi_{ml}^\Delta[\tau]}{\partial\eta_{ml}}~ \widetilde{\Phi}_l(\tau; \widetilde{\mathscr{F}}_T^m)  .
\end{align*}}

\textbf{Gradient of the mixture parameter.} For any $m\in \{1, \ldots, D\}$ and  for any $u \in \intervalleEntier{1}{N_{g_m}(T)} $, we have

{\small
\begin{align*}
\frac{\partial\bar{\mathcal{L}}_{\mathcal{G}}}{\partial \rho_{m}[u]}
     &= 2\Delta\sum_{i=1}^{D}H_i^1\mu_i   \sum_{\tau=1}^{L} \phi_{im}^{\Delta}[\tau] \bigg( \sum_{s=1}^{G}z_m[u] ~\mathbb{I}\{ u=s-\tau\} \bigg)\\&
     + 2 \Delta\sum_{i,k} H_i^1\sum_{\tau=1}^{L}\sum_{\tau'=1}^{L} \phi_{im}^\Delta[\tau] \phi_{ik}^\Delta[\tau'] \bigg(\sum_{s=1}^{G} \tilde{z}_k[s-\tau ']  z_m[u] ~\mathbb{I} \bigg\{u=s-\tau \bigg\}   \bigg)  \\&
     + \Delta \sum_{i=1}^{D}\sum_{\tau=1}^{L} \phi_{im}^{\Delta}[\tau]^2 \bigg( \sum_{s=1}^{G}  z_m[u] (z_m[u] - 2\tilde{z}_m[u])~\mathbb{I} \bigg\{u=s-\tau \bigg\}\bigg)\\&
     -2 \Bigg( -\tilde{\mu}_m \sum_{(\tilde{t}^m_n, \kappa^m_n) \in \widetilde{\mathscr{F}}_T^m)} f_i^0(\kappa^m_n)~\mathbb{I}\bigg\{u= \frac{\tilde{t}_n^m}{\Delta} \bigg \} + \mu_m \sum_{(t^m_n, \kappa^m_n) \in \widetilde{\mathscr{F}}_T^m)} f_i^1(\kappa^m_n)~\mathbb{I}\bigg\{u= \frac{\tilde{t}_n^m}{\Delta} \bigg \}
     \\&
     +\sum_{j=1}^{D} \sum_{\tau=1}^{L}\phi^{\Delta}_{mj}[\tau] \hspace{-0.2cm}\sum_{(\tilde{t}^m_n, \kappa_n^m) \in \mathscr F_T^m} \hspace{-0.2cm}f_m(\kappa^m_n)  \tilde{z}_j[u-\tau] \mathbb{I}\bigg\{u=\frac{\tilde{t}_n^m}{\Delta} \bigg \}\\&
    + \sum_{i=1}^{D} \sum_{\tau=1}^{L}\phi^{\Delta}_{im}[\tau] \sum_{(\tilde{t}^i_n, \kappa_n^i) \in \mathscr F_T^i} f_i(\kappa^i_n) \rho^{i}\bracks{u+\tau} z_{m}\bracks{u}\mathbb{I}\bigg\{u=\frac{\tilde{t}_n^i}{\Delta} - \tau \bigg \} \Bigg).
\end{align*}
}

\section{Additional Experiments}

\subsection{Sensitivity analysis of the alternate minimization parameter}
\label{subsec:batch}

The alternate minimization performed in UNHaP depends on a parameter $b$, the number of optimization steps done on the Hawkes parameters between each update of $\boldsymbol{\rho}$. It controls the trade-off between the number of gradients of the point process parameters and the latent variable $\boldsymbol{\rho}$. This part presents a sensitivity analysis of this parameter across several optimization iterations.

We conduct the experiment as follows. We simulate two univariate marked Hawkes processes with intensity functions defined as in (\ref{eq:sim}), the first one corresponding to the non-noisy setting with $\tilde{\mu}=0.1$ and the second one to the noisy setting with $\tilde{\mu}=1$. We set $T=1000$ for both settings. We set $\omega(\kappa)=\kappa$ and $f(\kappa) = 2\kappa ~\1[0\leq\kappa\leq 1]$ and the $g(\kappa) = \1[0\leq \kappa \leq 1]$. We set $\mu=0.8$, $\alpha=1.4$, imposing a high excitation phenomenon, and select $\phi^{\eta}$ to be a truncated Gaussian kernel with $W=1$ and $\eta=(m, \sigma)=(0.5, 0.1)$.

\begin{figure}[!ht]
    \centering
    \begin{tabular}{ccc}
    \includegraphics[trim=0cm 0cm 0cm 0cm,scale=0.35]{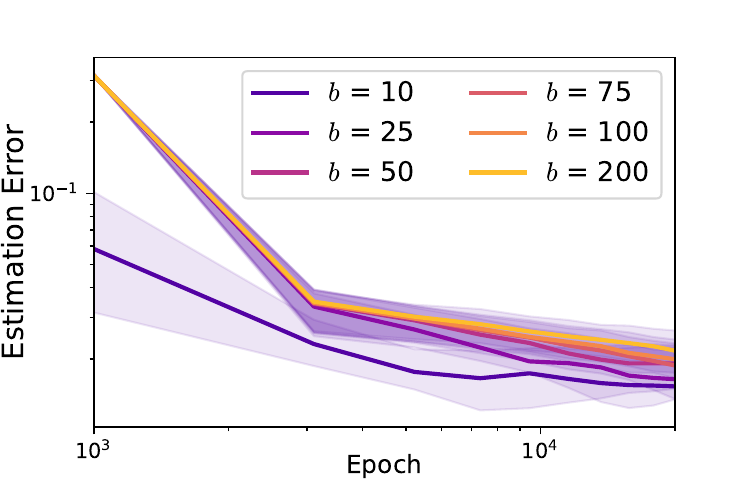} & \includegraphics[trim=0cm 0cm 0cm 0cm,scale=0.35]{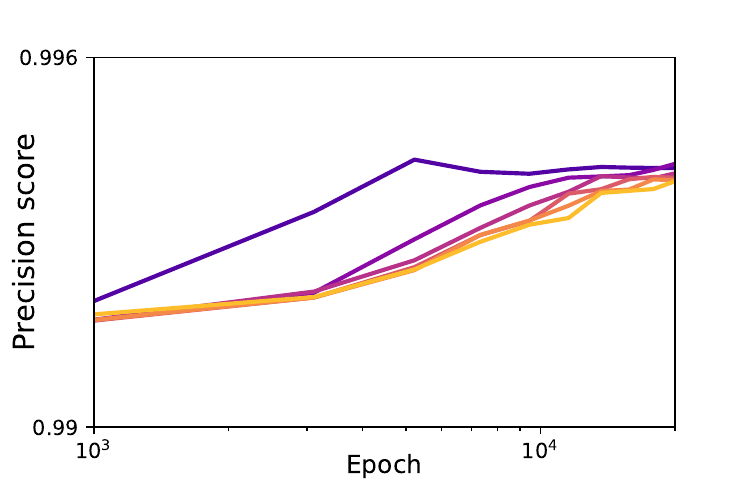} & \includegraphics[trim=0cm 0cm 0cm 0.cm,scale=0.35]{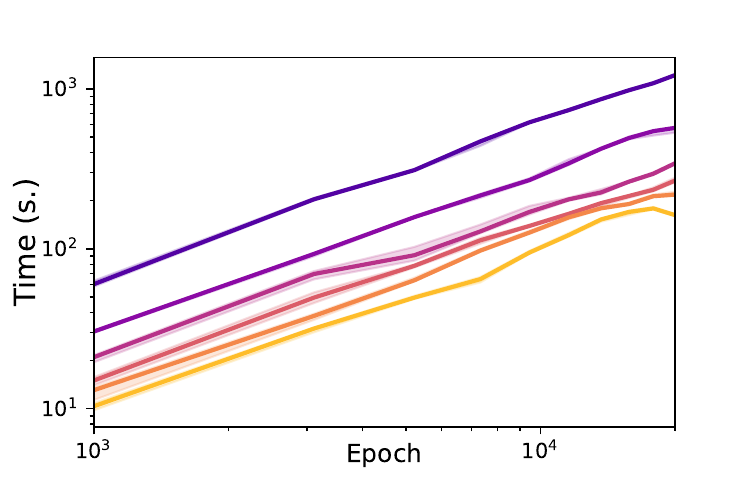} \\
    \includegraphics[trim=0cm 0cm 0cm 0.cm,scale=0.35]{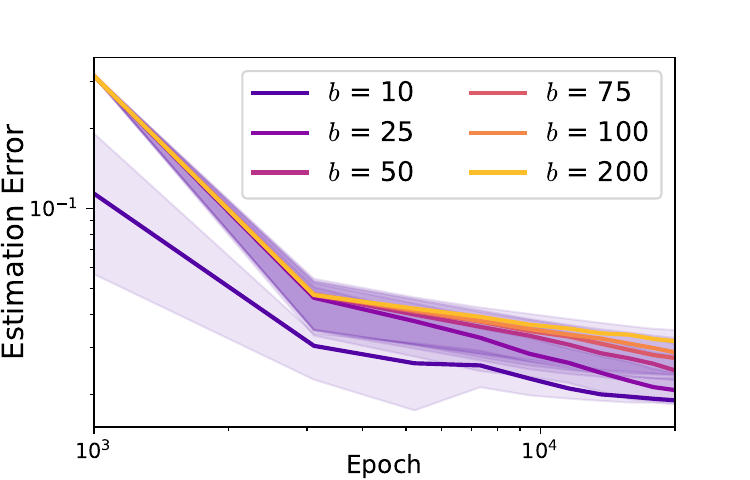} &\includegraphics[trim=0cm 0cm 0cm 0.cm,scale=0.35]{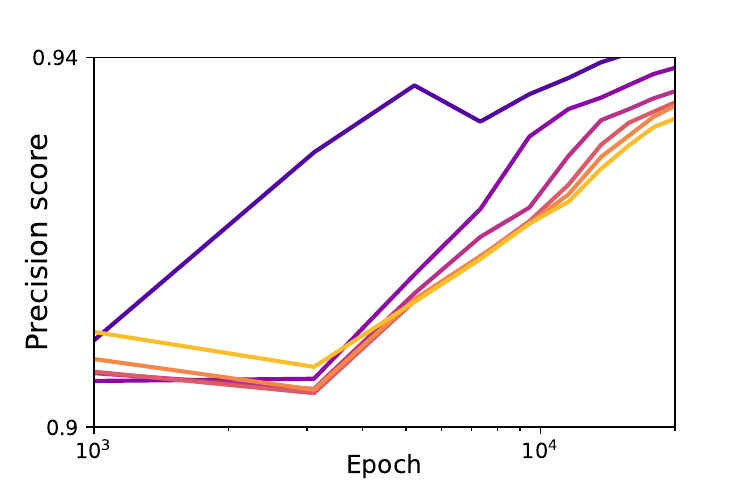} &  \includegraphics[trim=0cm 0cm 0cm 0.cm,scale=0.35]{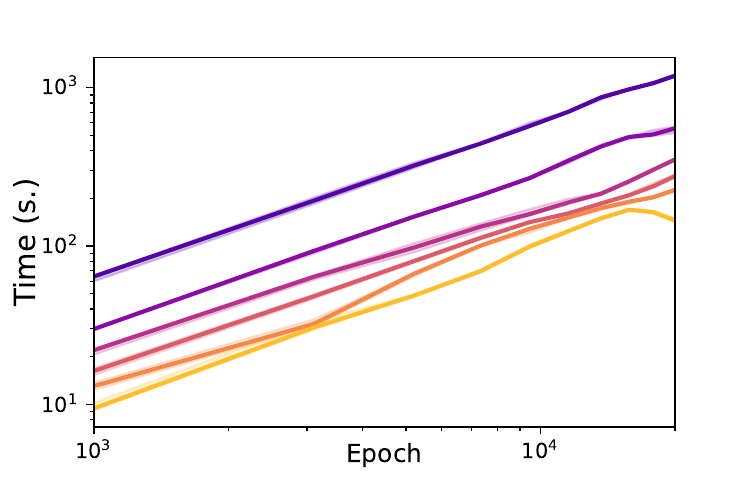}
    \end{tabular}
    \caption{Inference comparison regarding the batch size of Hawkes parameters gradients between each $\rho$ update. The error estimation on Hawkes parameters (left), the Precision score on the $\rho$ recovering (middle) and the associated computational time (right) are displayed for non-noisy (top) and noisy settings (bottom).}
    \label{fig:batch:appendix}
\end{figure}

We conduct inference on the intensity function of the underlying Hawkes processes using UNHaP with $\Delta=0.01$, $W=1$ and varying the value of $b$ in $\{10, 25, 50, 75, 100, 200 \}$. The median and the $25\%$-$75\%$ quantiles (over ten runs) of the estimation parameter are depicted in \autoref{fig:batch:appendix} (left) according to the number of iterations and the size of $b$. The median precision score (over ten runs) of the estimated $\hat{\boldsymbol{\rho}}$ recovery of the mixture structure parameter $\boldsymbol{\rho}$ is reported in \autoref{fig:batch:appendix} (middle). In both cases, and those for the two noisy and non-noisy settings, the size of $b$ reversely orders the accuracy at a computational cost; see \autoref{fig:batch:appendix} (right). However, the precision for each size $b$ is close to each other after 10000 iterations. Regarding the computational cost, we advise to select $b=200$ for UNHaP.

\subsection{Further experiments on the recovery of the mixture structure}

\begin{figure}[!ht]
    \centering
    \begin{tabular}{c|c}
    \includegraphics[trim=0cm 0cm 0cm 0cm,scale=0.45]{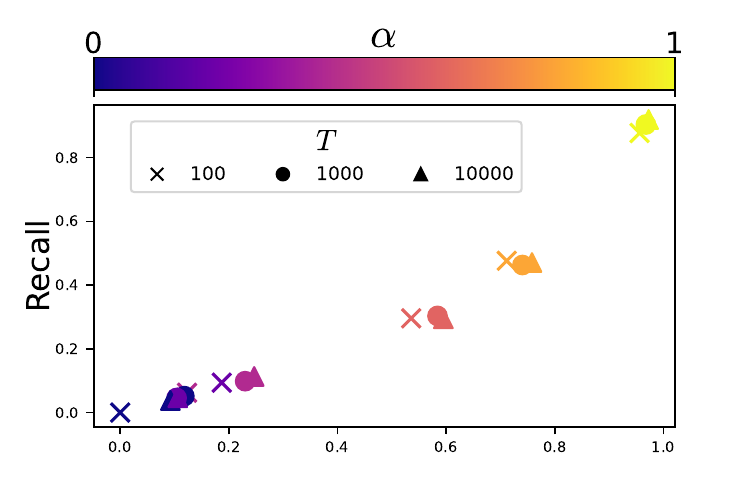} & \includegraphics[trim=0cm 0cm 0cm 0cm,scale=0.45]{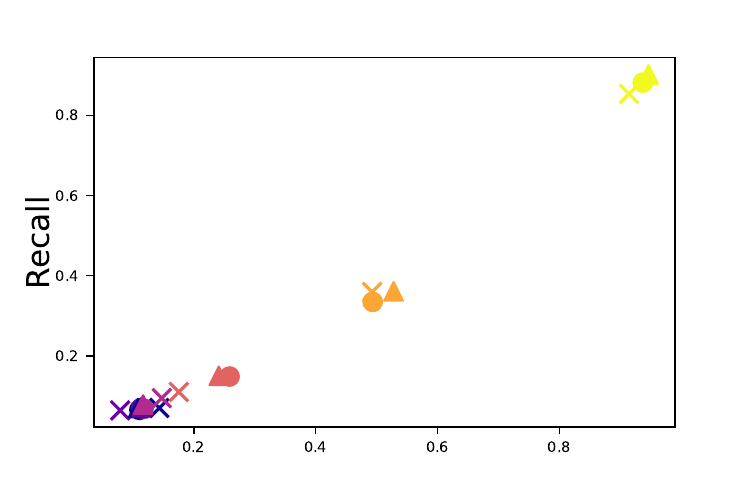}\\
    \includegraphics[trim=0cm 0cm 0cm 0.5cm,scale=0.45]{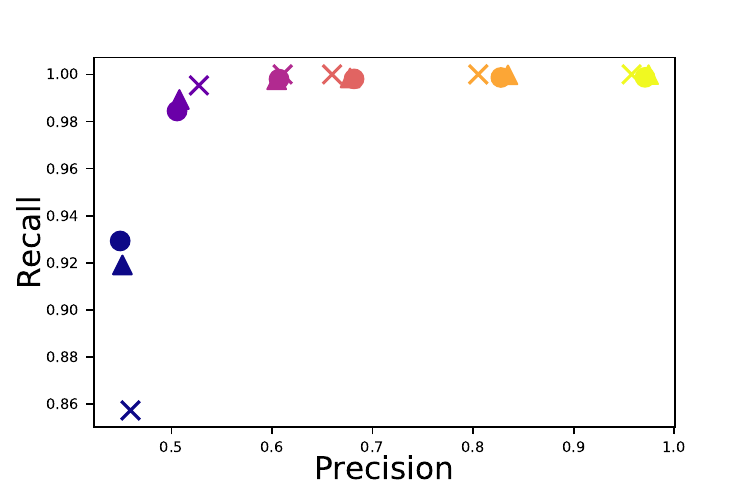} & \includegraphics[trim=0cm 0cm 0cm 0.5cm,scale=0.45]{scatter_rho_uniform.pdf}
    \end{tabular}
    \caption{Precision/Recall values for the estimation of $\boldsymbol{\rho}$ for different values of $T$ w.r.t. the auto-excitation parameter $\alpha$ with uniform (top) and linear (bottom) distributions on noisy marks for $\tilde{\mu}=0.5$ (left) and $\tilde{\mu}=1.$ (right).}
    \label{fig:rho:appendix}
\end{figure}

\autoref{fig:rho:appendix} displays the same experiment as in \autoref{subsec:adv} but with two different noise level $\tilde{\mu}=0.5$ and $\tilde{\mu}=1$. These additional experiments confirm and reinforce the claims made in the core paper regarding the recovery of the mixture structure of Hawkes processes polluted by Poisson processes.

\subsection{Moment Matching initialization}\label{subsec:mm:appendix}

This section investigates the advantages of using the Moment Matching initialization introduced in \autoref{subsec:mm} over the classical random ones.
The simulation study is conducted as follows. Relying on an Immigration-Birth algorithm \citep{moller2005perfect,moller2006approximate}, we simulate one-dimensional marked events in $[0, T] \times \mathcal{K}$ with $T=\{100, 1000, 10000 \}$ from the mixture process with the following intensity function

\begin{equation*}
    \lambda(t, \kappa; \boldsymbol{\theta}) = \bigg (\mu + \alpha \sum_{t_n<t} \omega(\kappa_n) \phi(t-t_n; \eta) \bigg ) f^1(\kappa) + \tilde{\mu}~f^0(\kappa),
\end{equation*}

where $\omega(\kappa)=\kappa$ and $f^1(\kappa) = 2\kappa ~\1[0\leq\kappa\leq 1]$. We define two settings of mark noise distribution: the ‘‘linear'' where $f^0(\kappa) = 2(1 -\kappa) ~\1[0\leq \kappa \leq 1]$ and the ‘‘uniform'' $f^0(\kappa) = \1[0\leq \kappa \leq 1]$. We set $\mu=0.8$ and $\alpha=1.4$ and $\tilde{\mu}=0.5$. Two excitation kernels $\phi (\cdot; \eta)$ are chosen. the first one is a truncated Gaussian, with $\eta=(m,\sigma)$, corresponding to
\begin{equation*}
    \phi(\cdot; \eta)  = \frac{1}{\sigma} \frac{\gamma\left(\frac{\cdot-m}{\sigma}\right)}{F\left(\frac{W-m}{\sigma}\right)-F\left(\frac{-m}{\sigma}\right)} \1[0\leq \cdot \leq W],
\end{equation*}
where $W$ is the kernel length and $\gamma$ (resp. $F$) is the probability density function (resp. cumulative distribution function) of the standard normal distribution.The second one is a raised cosine density defined as

\begin{equation*}
   \phi(\cdot; \eta)=\alpha \left[{1 + \cos \pars{\frac{\cdot - u}{\sigma}\pi - \pi}} \right] \1[u \leq \cdot \leq u+2\sigma],
\end{equation*}
with $\eta=(u, \sigma)$. In contrast to the truncated Gaussian, the support of this kernel directly depends on its parameters and may induce some instability in the optimization. For the truncated Gaussian kernel, we set $\eta=(m, \sigma)=(0.5, 0.1)$ while we set $\eta=(u, \sigma)=(0.4, 0.1)$ for the raised cosine.

We compute UNHaP with both Moments Matching and Random initialization (with $b=200$, $\Delta=0.01$) and report the error estimation (median over 10 runs)  between the true parameters $\boldsymbol{\theta}=\{\mu, \alpha, \eta \}$ and the estimated ones $\hat{\boldsymbol{\theta}}=\{\hat{\mu}, \hat{\alpha}, \hat{\eta} \}$, i.e.,
$||\hat{\boldsymbol{\theta}} -\boldsymbol{\theta} ||_2$.

 While the moment matching improves the convergence results of the parameter over the random initialization in the case of a raised cosine kernel, it behaves comparably in the case of a truncated Gaussian. This supports the average superiority of the moment matching over the random initialization and should be used consistently.

 \subsubsection{Benchmarking inference and computation time in an unmarked setting}
\label{sect:bench_supp}

The benchmark presented in \autoref{sect:bench} is done on simulated events with marks. However, the benchmarked methods do not account for marks, except for UNHaP, due to the scarcity of literature on marked point processes. To be exhaustive in our comparison, we present additional benchmarks of UNHaP, FaDIn, Tripp, and Neural Hawkes on unmarked events here.

{
\renewcommand{\arraystretch}{1.5} 
\setlength{\tabcolsep}{0.4cm}

\begin{table*}[!ht]
\begin{center}
{\scriptsize
\begin{tabular}{c|ccc|ccc || ccc}
& \multicolumn{6}{c}{NLL} &\\
\hline
& \multicolumn{3}{c}{Non-noisy} & \multicolumn{3}{c}{Noisy} &  \multicolumn{3}{c}{Computation time (s)} \\
\hline
$T$ & 100 & 500 & 1000 & 100 & 500 & 1000 & 100 & 500 & 1000\\
\hline
  UNHaP  &-0.18&\textbf{-1.7}& \textbf{-1.62}&  \textbf{1.18}& \textbf{-1.23} &\textbf{-1.20}  & 29& 31& 35 \\
  FaDIn & \textbf{-0.19}& \textbf{-1.7}& \textbf{-1.62} &1.2 &-1.18  & -1.17 & 3& 3& 3 \\
  Tripp & 2.9& -0.26&-0.98 & 5.4 & 2 & 1.71 & 19& 27& 31\\
  Neural Hawkes & 0.57& -1.27& -1.46 & 2.9&  1.87 & 1.66 & 20&149& 281 \\
\hline
\end{tabular}}
\end{center}
\caption{Median (over ten runs) Negative Log-Likelihood (NLL) \textbf{on unmarked events} in noisy and non-noisy settings for various models and various sizes of events sequence. Bold numbers correspond to the best results. Computation time associated with the non-noisy setting is also reported.}
\label{tab:bench_nomark}
\end{table*}

}

The events are simulated similarly to in \autoref{sect:bench}. We simulate a Marked Hawkes process. Its intensity function is defined as in (\ref{eq:sim}). We also simulate a Poisson Process for the noisy events. The marks are not taken into account here. Therefore, $\omega(\cdot)$, $f^1(\cdot)$ and $f^0(\cdot)$ are a Dirac function in one, $\delta_1(\cdot)$.  Similarly to the benchmark in \autoref{sect:bench}, we set $\mu=0.1$, $\alpha=1$, imposing a high excitation phenomenon, and select $\phi(\cdot; \eta)$ to be a truncated Gaussian kernel with width $W=1$ and parameters $\eta=(m, \sigma)=(0.5, 0.1)$. We benchmarked the methods on two noise settings: the non-noisy setting ($\tilde{\mu}=0.1$) and the noisy setting ($\tilde{\mu}=1$). Once the data is simulated, the inference and testing of the methods are done as developed in \autoref{sect:bench}.

The median Negative Log-Likelihood and computational time are shown in \autoref{tab:bench_nomark}. UNHaP demonstrates statistical superiority over all methods in a noisy environment while exhibiting comparable performance to FaDIn in a non-noisy context. This outcome aligns with expectations in a parametric approach when the utilized kernel belongs to the same family as the one used for event simulation. It is essential to highlight that these results stem from analyzing a single (long) data sequence, contributing to the subpar statistical performance of Neural Hawkes. It excels in scenarios involving numerous repetitions of short sequences due to the considerable number of parameters requiring inference. From a computational time standpoint, UNHaP performs similarly to Tripp, and significantly faster than Neural Hawkes. It is also slower than FaDIn, which is expected due to the alternate minimization scheme, which performs repeated parameter inference using a procedure similar to that of FaDIn. UNHaP offers an interesting alternative to existing methods in the context of unmarked noisy data at a reasonable computation cost.

 \subsection{Application to physiological data}
\label{sect:supp_physio}
 \subsubsection{ECG}

 Electrocardiograms (ECG) measure the electrical activity of the heart. They are the gold standard for observing heartbeats. Statistics derived from ECG, such as the heart rate (HR, average number of beats per minute) and the heart rate variability, are central in diagnosing heart-related health issues, like arrhythmia or atrial fibrillation \citep{shaffer2017overview}. These statistics require a robust estimate of the inter-beat interval duration.
To automatically measure the inter-beat interval, the first step is to accurately detect heartbeats \citep{berkaya2018survey}. This is usually done using knowledge-based methods based on analyses of slope, amplitude, and width of ECG waves \citep{Pan1985, elgendi2013fast, hamilton2002open}. However, raw ECG signals usually contain noise, which can lead to spurious event detection unrelated to the biological source of interest. These noisy events cause classical solvers to fail to recover the heart rate variability correctly. The usual route to circumvent this problem is handmade. It applies a post-processing step to the detected events, for instance, by thresholding them by amplitude or time-filtering them \citep{merdjanovska2022comprehensive}. The design of such a step is cumbersome, requires domain expertise, and does not generalize well. Indeed, ECG recordings often have considerable inter-individual variability, so it has no “one-fits-all" value.

The procedure we developed circumvents this problem by using the structure of the detected event location to remove spurious events. The underlying mixture model separates the data into events caused by the underlying Hawkes Process and events caused by noise.
In the following, we use UNHaP to post-process ECG events detected using CDL. Our results showcase that UNHaP filters out noisy events, and the obtained Hawkes process parameters are consistent with the biological ground.

\paragraph{Experimental Pipeline.}
Experiments are run on ECG data from the \textit{vitaldb} dataset \citep{lee2022vitaldb, goldberger2000physiobank}. Nineteen 5-minute long ECG slots were isolated among 7 patients and downsampled from 500 Hz to 200 Hz to reduce the computational cost. \autoref{fig:ecg} (A) shows a 3-second extract of an ECG slot. Each upward peak is a heartbeat. This succession of events is very regular and almost periodic. Hence, it is appropriate to model it with an MMHP and parameterize it with UNHaP. The downward peak at 1.5s is an example of an artifact. Below, we describe the event detection and UNHaP parameterization, done on each ECG slot separately.

We run a CDL algorithm to detect events using the Python library \href{https://alphacsc.github.io/}{\texttt{alphacsc}} \citep{latour2018multivariate}. Denote by $X$ an ECG slot, CDL decomposes it as a convolution between a dictionary of temporal atoms $D$ and a temporal activation vector $Z$: $X = Z*D + \varepsilon$. \autoref{fig:ecg} (B1) shows the learned temporal atom on ECG slot 1, and \autoref{fig:ecg} (B2) shows the learned activation vector $Z$ from ECG window in \autoref{fig:ecg} (A). There is  at least one non-zero activation for each beat. $Z$ could, therefore, be used as a proxy for event detection. In addition, noisy events are visible in $Z$: some are very close to beat activations, and some are caused by the ECG artifact at 1.5s. Handcrafted thresholding methods would typically be used here to remove noisy activations from $Z$. Instead, we process the raw activation vector $Z$, which is composed of sparse events, with our UNHaP solver with a truncated Gaussian kernel. The solver separates the heartbeat Hawkes Process from the noisy activations (\autoref{fig:ecg} (C2)) and estimates the inter-burst interval (\autoref{fig:ecg} (C1)). The mean (respectively the standard deviation) of the parameterized truncated Gaussian $\phi$ estimates the mean inter-beat interval (respectively the heart rate variability) on the ECG slot $X$.
With this example, we see that UNHaP successfully detects the structured events from the noisy ones, providing a good estimate of the inter-beat distribution.

\paragraph{Results.}
We compare the error made by several estimators of the inter-beat interval, including the developed pipeline with FaDIn \citep{staerman2023fadin} and UNHaP initialized with moment matching. We benchmark domain specific Python libraries \href{https://github.com/PGomes92/pyhrv/tree/master}{\texttt{pyHRV}} \citep{gomes_pgomes92pyhrv_2024} and \href{https://neuropsychology.github.io/NeuroKit/introduction.html#}{\texttt{Neurokit}} \citep{makowski_neurokit2_2021}, which are tailored to ECG.

The mean inter-beat interval given by these estimators is compared to the ground truth, which is given in the dataset. We average each estimator's absolute and relative absolute errors over the nineteen ECG slots, see \autoref{tab:bench_ecg}.
UNHaP, pyHRV, and Neurokit have equivalent performance. They all provide good heart rate estimates.
Another interesting finding is that while working with the same detected events, UNHaP vastly outperforms FaDIn. This proves the benefit of our mixture model to separate structured events --here quasi-periodic-- from spurious ones.


{\renewcommand{\arraystretch}{1.2} 
\begin{table}[h!]
\centering
\begin{tabular}{c|c|c|c|c}
& {\footnotesize CDL + UNHaP} & { \footnotesize CDL + FaDIn} & { \footnotesize pyHRV} & {\footnotesize Neurokit}\\
\hline
   {\small AE (beats/min)} & 0.61$\pm$0.93  & 15.44$\pm$19.39 & 0.57$\pm$0.14 & 0.45$\pm$0.07 \\
     {\small RAE}  & 0.009$\pm$0.012 & 0.192$\pm$0.227 & 0.008$\pm$0.002 & 0.006$\pm$0.001\\
\end{tabular}
\caption{Absolute Error (AE) and Relative Absolute Error (RAE) of the estimated average inter-beat interval (mean $\pm$ std) on the \textit{vitaldb} dataset.}
\label{tab:bench_ecg}
\vspace{-0.5cm}
\end{table}

\subsubsection{Gait}
\label{sect:gait}

The study of a person's manner of walking, or gait, is an important medical research field. Widespread pathologies, such as Parkinson's disease, arthritis, and strokes, are associated with an alteration of gait. Gait analysis is usually done by setting an inertial measurement unit to a patient's ankle and recording its vertical acceleration. These recordings can detect and infer essential features, such as steps, inter-step time intervals, and gait anomalies.
We applied CDL + UNHaP to gait inertial measurement unit recordings. Our pipeline detects steps and infers the inter-step time interval from raw gait inertial measurement unit data \citep{truong2019data}. We found that CDL+UNHaP performs at least as well as domain-specific methods.

\paragraph{Experimental pipeline}
\label{sect:gait_pipeline}
The experimental pipeline is the same as described in \autoref{sec:ecg}.
We run a CDL algorithm to detect steps using the Python library \href{https://alphacsc.github.io/}{\texttt{alphacsc}} \citep{latour2018multivariate}. The dictionary contains 1 atom of 1.5 seconds, and its loss is minimized with a regularization factor of 0.5. Detected events are then fed to the UNHaP solver. The Hawkes parameters are initialized with mean moment matching. The UNHaP gradient descent is done over 20,000 iterations, and the mixture parameter $\rho$ is updated every 1000 iterations.

\paragraph{Results}
\label{sect:gait_results}
We benchmarked our method to other estimators, similarly to \autoref{sec:ecg}
.
We compare the error made by several estimators of the inter-step interval, namely CDL + UNHaP developed in \autoref{sect:gait_pipeline} and CDL+FaDIn \citep{staerman2023fadin}. We compare these estimators with Template Matching \citep{oudre2018template}, a method specifically tailored for gait detection. Finally, the benchmark also includes the ECG Python libraries \href{https://github.com/PGomes92/pyhrv/tree/master}{\texttt{pyHRV}} \citep{gomes_pgomes92pyhrv_2024} and \href{https://neuropsychology.github.io/NeuroKit/introduction.html#}{\texttt{Neurokit}} \citep{makowski_neurokit2_2021}, which were benchmarked in \autoref{sec:ecg}.
The results are shown in \autoref{tab:bench_gait}. They highlight that as for ECG, UNHaP performs on par with Template Matching, which is state-of-the-art for gait detection. Additionally, contrary to previous Template Matching, UNHaP does not require pre-processing or application-specific tailoring. UNHaP performs much better on gait than the ECG methods pyHRV and Neurokit, illustrating its universality. We also stress that the proposed unmixing problem is critical to achieving good performance, as highlighted by the failure of CDL+FaDIn, which has no unmixing.

\begin{table}[h]
\centering
\begin{tabular}{c|c|c|c|c|c}

    & CDL + UNHaP & CDL + FaDIn & pyHRV & Neurokit & Template Matching
    \\\hline
MAE (seconds)
    & \textbf{0.04} & 1.2 & 0.67 & 0.57  & 0.07\\
\end{tabular}
\label{tab:bench_gait}
\caption{Median Absolute Error (MAE) of the estimated average inter-step interval on gait.}
\end{table}

\newpage

\end{document}

%% file: math_commands.tex
\usepackage{amsmath,amsfonts,bm}

\def\eqref#1{equation~\ref{#1}}

\def\floor#1{\left\lfloor #1 \right\rfloor}

\def\eps{{\epsilon}}

\DeclareMathAlphabet{\mathsfit}{\encodingdefault}{\sfdefault}{m}{sl}
\SetMathAlphabet{\mathsfit}{bold}{\encodingdefault}{\sfdefault}{bx}{n}

